\documentclass[reprint, superscriptaddress,
 amsmath,amssymb,
 aps,pra]{revtex4-2}

\usepackage{graphicx}
\usepackage[english]{babel}
\usepackage[utf8x]{inputenc}
\usepackage{dcolumn}
\usepackage[colorinlistoftodos]{todonotes}
\usepackage{xcolor}
\usepackage{mathtools}
\usepackage{bm}
\usepackage{hyperref}

\usepackage{psfrag}
\usepackage{indentfirst} 
\usepackage{url}
\usepackage{enumitem}
\usepackage{tensor}
\usepackage{braket}
\usepackage{bbm}
\usepackage{dsfont}
\usepackage{amsfonts,latexsym,cancel}
\usepackage{rawfonts}
\usepackage{pictexwd}
\usepackage{wrapfig}
\usepackage{subfigure} 
\usepackage{soul}

\DeclareMathOperator{\Tr}{Tr}

\usetikzlibrary{arrows}

\begin{document}

\preprint{}

\title{Classical analogs of the covariance matrix, purity, linear entropy, and von Neumann entropy}

\author{Bogar Díaz}
\email{bodiazj@math.uc3m.es}
\affiliation{Departamento de Matem\'aticas, Universidad Carlos III de Madrid, Avenida  de la Universidad 30, 28911 Legan\'es, Spain}
\affiliation{Departamento de F\'isica de Altas Energ\'ias, Instituto de Ciencias Nucleares, Universidad Nacional Aut\'onoma de M\'exico, Apartado Postal 70-543, Ciudad de M\'exico, 04510, M\'exico} 
\affiliation{Grupo de Teor\'ias de Campos y F\'isica Estad\'istica. Instituto Gregorio Mill\'an (UC3M), Unidad Asociada al Instituto de Estructura de la Materia, CSIC, Serrano 123, 28006 Madrid, Spain}
\author{Diego Gonz\'alez}
\email{diego.gonzalez@correo.nucleares.unam.mx}
\affiliation{Departamento de F\'isica de Altas Energ\'ias, Instituto de Ciencias Nucleares, Universidad Nacional Aut\'onoma de M\'exico, Apartado Postal 70-543, Ciudad de M\'exico, 04510, M\'exico} 
\author{Daniel Guti\'errez-Ruiz}
\email{daniel.gutierrez@correo.nucleares.unam.mx}
\affiliation{Departamento de F\'isica de Altas Energ\'ias, Instituto de Ciencias Nucleares, Universidad Nacional Aut\'onoma de M\'exico, Apartado Postal 70-543, Ciudad de M\'exico, 04510, M\'exico} 
\author{J. David Vergara}
\email{vergara@nucleares.unam.mx}
\affiliation{Departamento de F\'isica de Altas Energ\'ias, Instituto de Ciencias Nucleares, Universidad Nacional Aut\'onoma de M\'exico, Apartado Postal 70-543, Ciudad de M\'exico, 04510, M\'exico}

\begin{abstract}
    We obtain a classical analog of the quantum covariance matrix by performing its classical approximation for any continuous quantum state, and we illustrate this approach with the anharmonic oscillator. Using this classical covariance matrix, we propose classical analogs of the purity, linear quantum entropy, and von Neumann entropy for classical integrable systems, when the quantum counterpart of the system under consideration is in a Gaussian state. As is well known, this matrix completely characterizes the purity, linear quantum entropy, and von Neumann entropy for Gaussian states. These classical analogs can be interpreted as quantities that reveal how much information from the complete system remains in the considered subsystem. To illustrate our approach, we calculate these classical analogs for three coupled harmonic oscillators and two linearly coupled oscillators. We find that they exactly reproduce the results of their quantum counterparts. In this sense, it is remarkable that we can calculate these quantities from the classical viewpoint.
\end{abstract}

\maketitle

\section{Introduction}{\label{sec_introduction}}

Within the theory of quantum information, many tools help us describe the structure of the space of quantum states; in particular, it has been found that many structures of geometric type can be developed, and this branch of the theory is known as quantum information geometry. This branch aims to develop structures that allow us to establish a measure of distance between quantum states. Since quantum mechanics is a theory based on probability, the most logical idea to establish a geometric structure within this subject is to introduce a distance between probability distributions. In this way, Rokhlin \cite{Rokhlin_1967} and Rajski \cite{Rajski1961} introduced an information metric between two random variables; of course, the concept of statistical distance is entirely independent of quantum mechanics and can be defined in any probability space. However, using this metric, Schumacher \cite{Schumacher1991} was able to show the nonseparability of entangled states. On the other hand, the geometric structure of quantum mechanics has also borne many fruits, defining it through the parameter space using many quantities; among the most notable are the quantum fidelity \cite{Uhlmann1976}, the quantum metric tensor \cite{Provost1980}, the Berry phase \cite{Berry1984}, and the Loschmidt echo \cite{Gorin}. Furthermore, using a recent formulation \cite{ADV2017} it is possible to rewrite these geometric structures in terms of correlation functions of time-dependent operators. In contrast, within classical mechanics, the state space has a real differentiable manifold structure, and the observables correspond to the space of functions on the manifold. Despite this, there are several classical analogs of some tools found in quantum information geometry in the context of integrable systems, like the Hannay angle \cite{Hannay_1985} (see \cite{Chaturvedi1987,Zhang} for systems involving coherent states) and the classical analog of the quantum metric tensor \cite{ Gonzales2019, Alvarez2019, Gonzalez2020}. 

A fundamental property of quantum mechanics is entanglement, which plays a crucial role in quantum information and gives rise to quantities such as purity, linear quantum entropy, and von Neumann entropy. All these quantities are used to measure the degree of entanglement between quantum states, which is the quantum phenomenon par excellence \cite{Bell}. It is well known that quantum entanglement does not have a classical counterpart \cite{Paneru_2020}, however, some of the mentioned quantities do have a classic analog, such as linear quantum entropy \cite{Gong2003}. Furthermore, the quantum covariance matrix is a widely used tool for studying entanglement; see, for instance, \cite{Gittsovich2010} and references therein. An important feature of the quantum covariance matrix is that it fully characterizes Gaussian states, which play a crucial role in quantum information theory and many quantum optics experiments. For example, several quantum communication experiments use only Gaussian states \cite{Furusawa, Jeong}. In this line of thought, we wonder about the classical analogs of the quantum covariance matrix, purity, linear quantum entropy, and von Neumann entropy for classical integrable systems.

The purpose of this article is twofold. First, we  derive a classical analog of the quantum covariance matrix by performing its classical approximation in the context of integrable systems. Notably, the classical covariance matrix is able to produce the same results as its quantum counterpart, modulo the use of a quantization rule for the action variables. Second, using the classical covariance matrix, we propose classical analogs of the purity, linear quantum entropy, and von Neumann entropy for Gaussian states. Our classical analogs provide a measure of ``non-separability'' of the individual subsystems in phase space. In \cite{Collins-Popescu} it was suggested that a classical analogy of a quantum state $\psi_{ABE}$, in which the subsystems $A$ and $B$ are entangled with the environment $E$, is given by a probability distribution $P(X_A, X_B, X_E)$ where $X_A$, $X_B$, and  $X_E$ are random variables. In this paper, we elaborate more in the context of this analogy, considering that the role of the random variables is played by the classical phase-space variables expressed in terms of action-angle variables and the variances correspond to the correlation functions of the system.  An important feature of these classical analogs is that they can be calculated for any classical integrable system using classical tools only, with the advantage that they yield the same mathematical results as their quantum counterparts for Gaussian states, as demonstrated through the examples.   

The structure of the paper is the following. After this introduction, in Sec. \ref{sec:metric}, we first review the relationship between the quantum covariance matrix, symplectic matrix and quantum geometric tensor using the path integral formulation of quantum mechanics. Then, we obtain a classical covariance matrix for classical integrable systems, written in terms of action-angle variables, and establish its relation with the quantum covariance matrix. We show that this classical matrix yields the same results of its quantum counterpart for the ground state (which it is not Gaussian) of the anharmonic oscillator, modulo a quantization rule. In Sec. \ref{sec:puri},  we recall the definition of purity, linear quantum entropy, and von Neumann entropy. These functions are completely determined by the quantum covariance matrix of the system if it is in a Gaussian state. Then, taking this into account, in Sec. \ref{subsec:puri}, we introduce classical analogs of the purity, linear quantum entropy, and von Neumann entropy for classical integrable systems. These classical analogs are entirely determined by the classical covariance matrix of the classical system. In Sec. \ref{sec:exam}, we compute and compare our classical analogs of the purity, linear quantum entropy, and von Neumann entropy with their quantum counterparts for two examples: three coupled harmonic oscillators and two linearly coupled oscillators.  In  Sec. \ref{sec:conclu}, we give our conclusions and some comments. The paper ends with Appendixes \ref{apeA} and \ref{apenB}.

\section{Quantum and classical covariance matrices}\label{sec:metric}

The quantum covariance matrix $\boldsymbol{\sigma} = (\sigma_{\alpha \beta})$ of a quantum state $|m\rangle$ has entries given by
\begin{align}
\sigma_{\alpha \beta} := \frac{1}{2} \langle \hat{\bf{r}}_{\alpha} \hat{\bf{r}}_{\beta} + \hat{\bf{r}}_{\beta} \hat{\bf{r}}_{\alpha}  \rangle_{m}-\langle \hat{\bf{r}}_{\alpha}\rangle_{m} \langle \hat{\bf{r}}_{\beta} \rangle_{m}\,, \label{qumet}
\end{align}
where $\hat{\bf{r}}=(\hat{\bf{q}}_1,\dots,\hat{\bf{q}}_N,\hat{\bf{p}}_1,\dots,\hat{\bf{p}}_N)^T$ is a $2N$-dimensional column vector of position and momentum operators, $\alpha$ and $\beta$ run from 1 to 2$N$, and $ \langle \cdot \rangle_{m}=\langle m | \cdot  | m\rangle$ stands for the expectation value in the state $\ket{m}$. Throughout this paper, bold letters with a hat denote quantum operators. Before carrying out the classical approximation of the quantum covariance matrix~(\ref{qumet}), we want to show how it and the Fubini-Study metric are related, which can be performed using the path-integral formulation of quantum mechanics. Suppose that a Hamiltonian $H_i$ describes our system during the time interval $t \in (-\infty, 0)$, and assume that this Hamiltonian depends on a set of phase space coordinates $(q_a,p_a)$ with $a=1,\dots ,N$ and some parameters $\lambda_\iota$, with $\iota=1,\dots,M$. We shall consider that the full set of phase space coordinates and parameters is denoted by $z_A=(q_a,p_a, \lambda_\iota)$, in consequence, $A=1,\dots, 2N+M$. Furthermore, we assume that $H_i$ possesses a discrete and nondegenerate spectrum $E_n$.  Now, consider that at $t=0$, the system suffers a perturbation that modifies the Hamiltonian in the form
\begin{equation}\label{hamfin}
H_f = H_i + \mathcal{O}_A \delta z^A\,,
\end{equation}
where $ \mathcal{O}_A=\partial H_i / \partial z^A$  are deformation functions and  the system evolves from $(0,\infty)$  with this new Hamiltonian $H_f$. Notice that here we are considering more general variations than those given in \cite{ADV2017} since we allow variations of the phase space variables $(q_a,p_a)$. However, these variations could only be translations in such a way that we can extract them from the quantum averages. Now, following a procedure analogous to that presented in \cite{ADV2017}, we can arrive at a \textit{generalized} quantum geometric tensor for the $m-$th state
\begin{align} 
G_{AB}^{(m)}  =   -\frac{1}{\hbar^2}\int_{-\infty}^0 \!\!\! \!\!\! \mathrm{d}t_1 \!\! \int_0^{\infty} \!\!\! \!\!\! \mathrm{d}t_2 &\left[  \left<  \boldsymbol{\hat{\mathcal{O}}}_A(t_1) \boldsymbol{\hat{\mathcal{O}}}_B(t_2)\right>_m \right. \nonumber \\
&\left. - \left<\boldsymbol{\hat{\mathcal{O}}}_A(t_1) \right>_m \left< \boldsymbol{\hat{\mathcal{O}}}_B(t_2)\right>_m \right]\,.\label{eq:defQGT}
\end{align}
The real part of this tensor gives the Fubini-Study metric, $g_{AB}^{(m)} = \textbf{Re}\, G_{AB}^{(m)}$, and its imaginary part is related to the (generalized) Berry curvature, $F_{AB}^{(m)} = -2\,\textbf{Im}\, G_{AB}^{(m)}$. An alternative expression for \eqref{eq:defQGT} is given in \cite{Abe1993}.
In this way, we have full expressions for the quantum metric tensor and the Berry curvature for the $m-$th quantum state and defined for arbitrary variations in the parameter-space and phase-space translations.

The phase space part of $g_{AB}^{(m)}$ is related to the quantum covariance matrix in the following way
\begin{subequations} \label{rfscm}
\begin{align}
g_{q_{a}q_{b}}^{(m)}&=\frac{1}{\hbar^2}\left( \frac{1}{2}\langle \hat{\bf{p}}_{a} \hat{\bf{p}}_{b}+\hat{\bf{p}}_{b} \hat{\bf{p}}_{a} \rangle_{m} - \langle \hat{\bf{p}}_{a} \rangle_{m} \langle \hat{\bf{p}}_{b} \rangle_{m} \right)\,,\\
g_{q_{a}p_{b}}^{(m)}&=\frac{-1}{\hbar^2}\left( \frac{1}{2}\langle \hat{\bf{p}}_{a} \hat{\bf{q}}_{b}+\hat{\bf{q}}_{b} \hat{\bf{p}}_{a} \rangle_{m} - \langle \hat{\bf{p}}_{a} \rangle_{m} \langle \hat{\bf{q}}_{b} \rangle_{m} \right)\,,\\
g_{p_{a}p_{b}}^{(m)}&=\frac{1}{\hbar^2}\left( \frac{1}{2}\langle \hat{\bf{q}}_{a} \hat{\bf{q}}_{b}+\hat{\bf{q}}_{b} \hat{\bf{q}}_{a} \rangle_{m} - \langle \hat{\bf{q}}_{a} \rangle_{m} \langle \hat{\bf{q}}_{b} \rangle_{m} \right)\,.
\end{align}
\end{subequations}
The proof of these equations is given in Appendix \ref{apeA}. In this way, we observe that the Fubini-Study metric for the phase space variables reduces to the quantum covariance matrix (\ref{qumet}), modulo the $1/\hbar^2$ factor and a sign in the components $g_{q_{a}p_{b}}^{(m)}$. As a closing remark, it is worth noting that the phase space part of the generalized Berry curvature, $F_{\alpha \beta}^{(m)}$, is related to the symplectic matrix $\Omega_{\alpha \beta}:=-i [\hat{\bf{r}}_{\alpha},\hat{\bf{r}}_{\beta}]$ as
\begin{equation}\label{symplecticM}
    F_{\alpha \beta}^{(m)}=-\frac{1}{\hbar^2} \Omega_{\alpha \beta}\,.
\end{equation}
Thus, the generalized quantum geometric tensor (\ref{eq:defQGT}) contains both the quantum covariance matrix and the symplectic matrix. Notice that the quantum geometric tensor associated to the parameters admits a classical analog. Then, it is natural to look for a classical analog of the phase space part of \eqref{eq:defQGT}.

\subsection{Classical approximation of the quantum covariance matrix}\label{subsec:cmetric}

In this section, we consider classical integrable systems for which the action-angle variables, $I=\{I_a\}$ and $\varphi=\{\varphi_a\}$, exist and prove that the quantum covariance matrix \eqref{qumet} reduces to the ``classical'' covariance matrix $\sigma^{\mathrm{cl}}= \left(\sigma^{\mathrm{cl}}_{\alpha \beta} \right)$, with matrix elements
\begin{equation}
\sigma^{\mathrm{cl}}_{\alpha \beta} :=  \langle r_{\alpha} r_{\beta} \rangle_{\mathrm{cl}}-\langle r_{\alpha} \rangle_{\mathrm{cl}} \langle  r_{\beta} \rangle_{\mathrm{cl}} \,, \label{eq:sigmac}
\end{equation}
in the classical approximation, i.e., when $\hbar \to 0$, $m \to \infty$ such that $m \hbar$ is constant and equal to a particular torus $I_m$ corresponding to a quantum 
state $\left| m\right>$. In \eqref{eq:sigmac},  $r=(q_1,\dots,q_N,p_1,\dots,p_N)^T$ is a $2N$-dimensional phase-space column vector and we define
\begin{equation}\label{classAvg}
\langle f \rangle_{\mathrm{cl}}:= \frac{1}{\left( 2 \pi\right)^N}  \int \!\!\! \cdots \!\!\! \int_{0}^{2 \pi} \!\!\! \mathrm{d}^N \varphi \, f  \,,
\end{equation}
which denotes the classical average of a function $f=f(\varphi,I)$.  Here, we write $\mathrm{d}^N \varphi = {\rm d} \varphi_1 \, \dotsi \, {\rm d} \varphi_N$. Notice that \eqref{eq:sigmac}, as well as \eqref{qumet}, is symmetric $(\sigma^{\mathrm{cl}}_{\alpha \beta}=\sigma^{\mathrm{cl}}_{\beta \alpha})$ and positive semidefinite. In Appendix \ref{apenB}, we provide another perspective on the classical average \eqref{classAvg}.

The proof can be done straightforwardly by using the Wigner-function formalism \cite{Wigner1932,Hillery1984, Case2008, Gonzalez2020}. In this approach, the expectation value of an operator $\hat{\bf{O}}(\hat{\bf{q}},\hat{\bf{p}})$ can be written as
\begin{eqnarray}\label{WignerAvg}
	\langle \hat{\bf{O}} \rangle_m =
	\int_{-\infty}^{\infty} \!\!\! {\rm d}^N q \, {\rm d}^Np\, W_m \, \mathcal{O} \,,
\end{eqnarray}
where $W_m$ is the Wigner function and $\mathcal{O}$ is the Weyl transform of $\hat{\bf{O}}$, which are respectively given by
\begin{subequations}
\begin{align}
W_m(q,p) &= \frac{1}{(2\pi \hbar)^N} \int_{-\infty}^{\infty} \!\!\! {\rm d}^N z \,  e^{ -\frac{{\rm i}p\cdot z}{\hbar}} \psi_{m}(q+\tfrac{z}{2})\psi^*_{m}(q-\tfrac{z}{2})  \,, \label{eq:wf}\\
\mathcal{O}(q,p)&=\int_{-\infty}^{\infty} \!\!\! {\rm d}^N z \, e^{ -\frac{{\rm i}p\cdot z}{\hbar}} \bra{q+\tfrac{z}{2}} \hat{\bf{O}}(\hat{\bf{q}},\hat{\bf{p}})\ket{q-\tfrac{z}{2}}\,.
\end{align}
\end{subequations}
Here, the variables $q=\{q_a\}$ and $p=\{p_a\}$ are the eigenvalues of the operators $\hat{\bf{q}}$ and $\hat{\bf{p}}$, respectively. Also, we have written $p\cdot z =\sum_{a=1}^{N} p_a z_a$, and $\psi_{m}(q+\tfrac{z}{2})=\braket{q+\tfrac{z}{2}|  m }$, $\psi^*_{m}(q-\tfrac{z}{2})=\braket{m | q-\tfrac{z}{2}}$. Some useful results are that if $\hat{\bf{F}}= F(\hat{\bf{q}})$, $\hat{\bf{G}}= G(\hat{\bf{p}})$, and $\hat{\bf{K}}=\frac{1}{2} \left(  \hat{\bf{q}}_a \hat{\bf{p}}_b + \hat{\bf{p}}_b \hat{\bf{q}}_a  \right)$, then their Weyl transform satisfy $\mathcal{F}=  F(q)$, $\mathcal{G}= G (p)$, and $\mathcal{K}= q_a p_b $, respectively \cite{Case2008}.

Let us now consider a quantum system whose classical motion is integrable. In the classical limit (which we denote by $\simeq$), the Wigner function $W_m(q,p)$ becomes a delta function on the torus $I_m$ associated with the quantum state $\left| m\right>$~\cite{Berry1977}. More precisely, it is given by
\begin{equation}
W_m (q,p) \simeq \frac{1}{(2 \pi)^N} \delta \left( I(q,p)-I_m \right) \,. \label{eq:wfca}
\end{equation}
Using \eqref{eq:wfca} together with \eqref{WignerAvg} and \eqref{classAvg}, we can show that
\begin{align}
\langle \hat{\bf{q}}_a \rangle_m & \simeq \int_{-\infty}^{\infty} \!\!\!  {\rm d}^N q \, {\rm d}^Np\, \frac{1}{(2 \pi)^N} \delta \left( I(q,p)-I_m \right) q_a\nonumber\\
&=\frac{1}{\left( 2 \pi\right)^N} \int_{0}^{\infty} \!\!\! \mathrm{d}^N I    \int_{0}^{2 \pi} \!\!\! \mathrm{d}^N \varphi \,  \delta \left( I-I_m \right) q_a(I, \varphi)\nonumber\\
&=  \frac{1}{\left( 2 \pi\right)^N}    \int_{0}^{2 \pi} \!\!\! \mathrm{d}^N \varphi \,   q_a(I_m, \varphi)\nonumber\\
&= \langle q_a \rangle_{\mathrm{cl}}\,. \label{weylq}
\end{align}
By following an analogous procedure, it is possible to prove the following relations
\begin{subequations} \label{eq:qce}
\begin{align}
\langle \hat{\bf{q}}_a  \hat{\bf{q}}_b \rangle_m  & \simeq \langle q_a q_b \rangle_{\mathrm{cl}}\,,\label{weylq1}\\
\langle \hat{\bf{p}}_a  \rangle_m                & \simeq \langle p_a \rangle_{\mathrm{cl}} \,,\\
\langle \hat{\bf{p}}_a  \hat{\bf{p}}_b \rangle_m  & \simeq \langle p_a p_b \rangle_{\mathrm{cl}}\,,\\
\frac{1}{2} \langle \hat{\bf{q}}_a \hat{\bf{p}}_b + \hat{\bf{p}}_b \hat{\bf{q}}_a   \rangle_m  &  \simeq \langle q_a p_b  \rangle_{\mathrm{cl}}\,, 
\end{align}
\end{subequations}
Plugging \eqref{weylq} and \eqref{eq:qce} into \eqref{qumet}, it follows that $\sigma_{\alpha \beta} \simeq \sigma^{\mathrm{cl}}_{\alpha \beta}$, and hence
\begin{equation}
 \boldsymbol{\sigma} \simeq \sigma^{\mathrm{cl}}\, , \label{eq:equicovma}
\end{equation}
which completes the proof. Then, we show that, for a quantum system whose classical counterpart is integrable, the classical analog of $\boldsymbol{\sigma}$ is given by $\sigma^{\mathrm{cl}}$. This means that we can get a first glimpse of the quantum covariance matrix~$\boldsymbol{\sigma}$ from the classical framework by using~$\sigma^{\mathrm{cl}}$. 

To complete the scheme, we introduce the classical analog of the symplectic matrix \eqref{symplecticM}. In this case, to order $\hbar$ we now obtain 
\begin{align}
\frac{1}{i\hbar}\langle \hat{\bf{q}}_a  \hat{\bf{p}}_b - \hat{\bf{p}}_b  \hat{\bf{q}}_a  \rangle_m&=\frac{1}{i\hbar} \langle q_a *p_b - p_b * q_a  \rangle_m \nonumber\\
&= \langle \{ q_a, p_b\} \rangle_{\mathrm{cl}}\,,
\end{align}
where to take the classical limit, we first transform all the operators to functions using the phase-space formalism of quantum mechanics and the Moyal product \cite{Zachos}, and $\{q_a,p_b\}$ corresponds to the Poisson bracket.
On the other hand, since the classical covariance matrix is positive-semidefinite it follows that the corresponding uncertainty relation between the functions $r_\alpha$ and  $r_\beta$ is 
\begin{align}
\sigma^{\mathrm{cl}}_{\alpha \alpha} \sigma^{\mathrm{cl}}_{\beta \beta}   \geq 0\,. \label{Diego}
\end{align}
See \cite{Huang2011} for a general classical statistical uncertainty relation.

In the next sections, we exploit the consequences of  \eqref{eq:equicovma} and propose, for integrable systems, classical analogs of some quantum quantities that can be written as a function of the quantum covariance matrix.

\subsection{Classical and quantum covariance matrices of the quartic anharmonic oscillator}

Here, we illustrate the computation of the classical covariance matrix for the quartic anharmonic oscillator and compare it with its quantum counterpart. We consider the Hamiltonian
\begin{equation}\label{HQuartic}
H=\frac{1}{2m}p^2+\frac{m\omega_0^2}{2}q^2+\frac{\lambda}{4!}q^4\,,
\end{equation}
with $\lambda>0$ but $\lambda \ll 1$ so that we can treat the system perturbatively.

We begin by splitting the Hamiltonian into two pieces: the unperturbed exactly-solvable Hamiltonian $H_0=\frac{1}{2m}p^2+\frac{m\omega_0^2}{2}q^2$, and the perturbation $H_1=\frac{1}{4!}q^4$. In this way, \eqref{HQuartic} reads as
\begin{equation}
H=H_0+\lambda H_1\,.
\end{equation}
The action-angle variables $(\varphi_0,I_0)$ of $H_0$ are
\begin{subequations}\label{ActionAngleQuartic}
\begin{align}
q &= \sqrt{\frac{2I_0}{m\omega_0}}\sin \varphi_0\,, \\
p &= \sqrt{2I_0 m\omega_0}\cos \varphi_0\,.
\end{align}
\end{subequations}
To find the action-angle variables $(\varphi,I)$ of the complete Hamiltonian~\eqref{HQuartic}, we need the generating function of the canonical transformation $(\varphi_0,I_0)\to(\varphi,I)$ which we call $W$. This function depends on the mixed pair $(\varphi_0,I)$ and is expressed as a power series in $\lambda$:
\begin{equation}
W=\varphi_0 I+\lambda W_1+\lambda^2 W_2+\dots \,,
\end{equation}
where the functions $W$s are computed through canonical perturbation theory~\cite{Dittrich}. This in turn allows us to obtain $I_0$ and $\varphi$ as
\begin{subequations}\label{PowerSeriesQuartic}
\begin{align}
I_0 &= \frac{\partial W}{\partial \varphi_0} = I+\lambda\frac{\partial W_1}{\partial \varphi_0}+\lambda^2\frac{\partial W_2}{\partial \varphi_0}+\dots \label{PowerSeriesQuartic_a}\,, \\
\varphi &= \frac{\partial W}{\partial I} = \varphi_0+\lambda\frac{\partial W_1}{\partial I}+\lambda^2\frac{\partial W_2}{\partial I}+\dots \,.\label{PowerSeriesQuartic_b}
\end{align}
\end{subequations}
The substitution of~\eqref{PowerSeriesQuartic_a} into~\eqref{ActionAngleQuartic} and a further expansion in $\lambda$ gives us the required elements to compute the classical covariance matrix. All that remains is to change the measure of the classical average~\eqref{classAvg} as $\mathrm{d}\varphi= \left( \partial \varphi/\partial\varphi_0 \right)\mathrm{d}\varphi_0$ using~\eqref{PowerSeriesQuartic_b} to facilitate the integration. The averages $\langle q\rangle_{\text{cl}}$ and $\langle p\rangle_{\text{cl}}$ vanish, and the components of resulting classical covariance matrix to order $\lambda^2$ are
\begin{subequations}\label{ClassCovQuartic}
\begin{align}
\sigma^{\text{cl}}_{11} &= \langle q^2\rangle_{\text{cl}} = \frac{I}{m\omega_0}-\frac{\lambda I^2}{8m^3\omega_0^4}+\frac{85\lambda^2 I^3}{2304m^5\omega_0^7}+\dots \,,\\
\sigma^{\text{cl}}_{12} &= \langle qp\rangle_{\text{cl}} = 0 \,, \\
\sigma^{\text{cl}}_{22} &= \langle p^2\rangle_{\text{cl}} =  m  \omega_0 I+\frac{\lambda I^2}{8m\omega_0^2}-\frac{17\lambda^2 I^3}{768m^3\omega_0^5}+\dots \,.
\end{align}
\end{subequations}

Now, on the quantum side, we compute the quantum covariance matrix for the ground state only, which already departs from a Gaussian state. To this end, we use the perturbative procedure of nonlinearization to obtain the wave function as a power series in $\lambda$~\cite{Turbiner1984}. Once the wave function is obtained, we can compute the required expectation values, finding that $\langle\hat{\bf{q}}\rangle_0$ and $\langle\hat{\bf{p}}\rangle_0$ vanish. The quantum covariance matrix to order $\lambda^2$ turns out to be
\begin{subequations}\label{QuantCovQuartic}
\begin{align}
\sigma_{11} &= \langle \hat{\bf{q}}^2\rangle_0 = \frac{\hbar}{2m\omega_0}-\frac{\lambda\hbar^2}{16m^3\omega_0^4}+\frac{35\lambda^2\hbar^3}{1536m^5\omega_0^7}+\dots \,,\\
\sigma_{12} &= \frac{1}{2}\langle \hat{\bf{q}}\hat{\bf{p}}+\hat{\bf{p}}\hat{\bf{q}}\rangle_0 = 0 \,, \\
\sigma_{22} &= \langle \hat{\bf{p}}^2\rangle_0 = \frac{m\omega_0\hbar}{2}+\frac{\lambda\hbar^2}{16m\omega_0^2}-\frac{7\lambda^2\hbar^3}{512m^3\omega_0^5}+\dots \,.
\end{align}
\end{subequations}

To compare~\eqref{ClassCovQuartic} with~\eqref{QuantCovQuartic}, we need to establish a quantization prescription for the action variable. A first and simple approach, although not exact, is to use the quantization rule of the harmonic oscillator for the ground state $I=\hbar/2$; this yields important differences between the coefficients of the terms in both series. A second approach is to use different quantizations for different powers of the action variable, analogous to what was done in \cite{Gonzales2019} and \cite{Alvarez2019}. In this case, the rules are $I=\hbar/2$, $I^2=\hbar^2/2$, and $I^3=21\hbar^3/34$, which substituted into~\eqref{ClassCovQuartic} give rise to~\eqref{QuantCovQuartic} and are precisely the same rules that reproduce the energy series~\cite{Flugge,Thesis2021}. This shows that, even for this non-quadratic system, the classical covariance matrix leads to the same results as its quantum counterpart modulo a quantization rule.

In the light of the classical uncertainty relation \eqref{Diego}, we draw our attention to the product of the variances $(\Delta r_\alpha)^2=\langle \left(  r_\alpha- \langle r_\alpha \rangle \right)^2 \rangle= \langle r^2_\alpha \rangle- \langle r_\alpha \rangle^2= \sigma^{\mathrm{cl}}_{\alpha \alpha} $ of $q$ and $p$. Using \eqref{ClassCovQuartic} we obtain, in the classical case,
\begin{equation}
(\Delta q)^2(\Delta p)^2 = I^2-\frac{\lambda^2I^4}{1152m^4\omega_0^6}+\dots \,,
\end{equation}
and using \eqref{QuantCovQuartic}, the result in the quantum case is
\begin{equation}
(\Delta\hat{\bf{q}})^2(\Delta\hat{\bf{p}})^2 = \frac{\hbar^2}{4}+\frac{\lambda^2\hbar^4}{1536m^4\omega_0^6}+\dots \,.
\end{equation}
Clearly, the zeroth order in $\lambda$ corresponds to the well-known result of the harmonic oscillator.

\begin{figure}[ht]
\includegraphics[width=.4\textwidth]{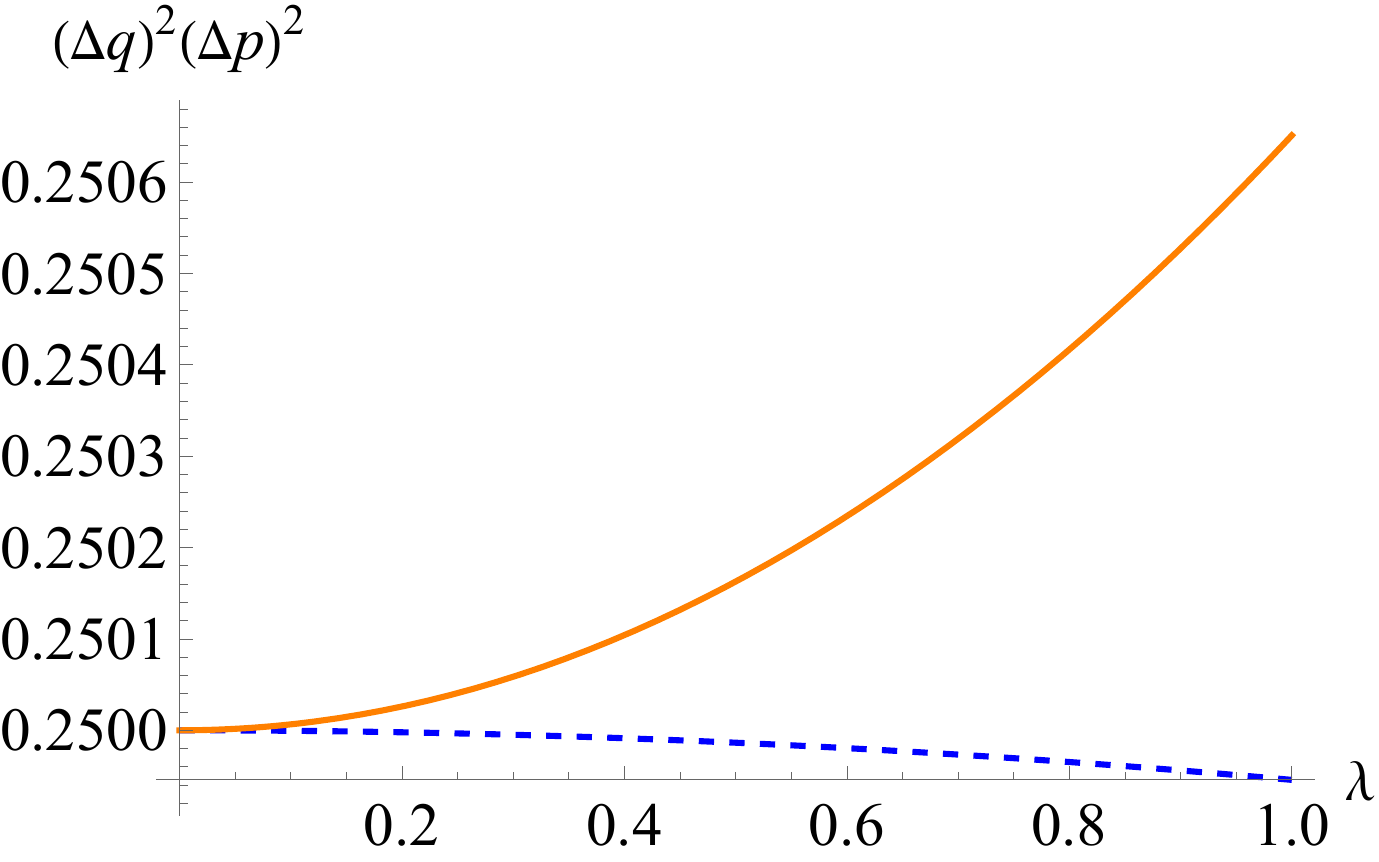}
\caption{Plot of $(\Delta q)^2(\Delta p)^2$ as a function of $\lambda$ fixing $m=1$, $\omega_0=1$, $\hbar=1$, and $I=1/2$. The dashed blue curve corresponds to the classical result and the solid orange corresponds to the quantum result.}
\label{Figure_var}
\end{figure}

In Fig.~\ref{Figure_var}, we show the plots of both functions and see that the curve of the quantum result grows with $\lambda$, as opposed to the classical result. This illustrates the fact that the product of variances in the quantum case must be greater than or equal to $\hbar^2/4$, which is a stronger constraint than the non-negativity of the classical case.

\section{Purity, linear quantum entropy, and von Neumann entropy} \label{sec:puri}

The purity $\mu$ of a normalized quantum state, described by a density operator $\hat{\rho}$, is defined as~\cite{deGosson2006, diosi2011}
\begin{equation}
\mu \left( \hat{\rho}\right)= \Tr \hat{\rho}^2\,.   \label{eq:pur}
\end{equation}
Notice that for pure states $\hat{\rho}^2 = \hat{\rho}$, and then $\mu$ takes a maximum value
of $1$. On the other hand, for mixed states $\hat{\rho}^2 \neq \hat{\rho}$ and thus $0< \mu < 1$. The purity is related to the linear quantum entropy $S_L$ by
\begin{align}
S_L \left( \hat{\rho}\right) = 1-\mu\left( \hat{\rho}\right)\,, \label{eq:lS} 
\end{align}
which provides a measure of the degree of mixedness of a quantum state.
 
So far, we consider general quantum states. In what follows, we restrict our analysis to Gaussian states, which are well known for being fully characterized by the quantum covariance matrix and the first moments of the canonical operators \cite{diosi2011, deGosson2006}. The motivation is that, for these states, the purity depends only on the quantum covariance matrix \cite{Dodonov_2002, Paris2003, deGosson2006, Golubeva2014, serafini2017, deGosson2019}. In fact, considering an $n$-mode Gaussian state with quantum covariance matrix $\boldsymbol{\sigma}_{(n)}$ ($n$ denotes the degrees of freedom of the subsystem formed by the particles $a_1,a_2,...,a_n$  of the system of $N$ degrees of freedom), the purity \eqref{eq:pur} takes the simple form
\begin{equation}
\mu \left(a_1,a_2,...,a_n\right)= \left( \frac{\hbar}{2}\right)^n \frac{1}{\sqrt{\det \boldsymbol{\sigma}_{(n)} }} \,. \label{eq:qpu}
\end{equation}
Here, the subscript $(n)$ represents $(a_1,a_2,...,a_n)$, for a compact notation. We must notice that written in this way, the purity $\mu$ has a statistical interpretation in terms of the quantum covariance matrix or the Fubini-Study metric, which tells us how much information from the complete system remains in the subsystem considered. This fact will allow us to give a classical interpretation of the quantum purity, as we will see in the following section.

The name ``linear quantum entropy'' of $S_L$ follows from the fact that it corresponds to a lower approximation of the von Neumann entropy $S$, which for a quantum state $\hat{\rho}$ is defined by
\begin{equation}
S \left(\hat{\rho}\right)= -\Tr \left( \hat{\rho}  \ln \hat{\rho}  \right)\,. \label{eq:entro}
\end{equation}
The von Neumann entropy is zero for pure states and serves as a measure of the mixedness of the quantum state \cite{diosi2011}. Notably, for an $n$-mode Gaussian state with covariance matrix $\boldsymbol{\sigma}_{(n)}$, this entropy can be written as \cite{Agarwal1971, Holevo1999, deGosson2006, serafini2017, Demarie2018}
\begin{equation}\label{eq:Sred}
    S \left( a_1,a_2,...,a_n\right) = \sum_{k=1}^{n} \mathcal{S} (\nu_k)\,, 
\end{equation}
with
\begin{equation}
   \mathcal{S}(\nu_k)=  \left( \nu_k+\frac{1}{2} \right)\ln \left( \nu_k+\frac{1}{2} \right)-\left( \nu_k-\frac{1}{2} \right) \ln \left( \nu_k-\frac{1}{2} \right)\,, 
\end{equation}
where $\nu_k$ are the symplectic eigenvalues of $\boldsymbol{\sigma}_{(n)} / \hbar$, i.e., they are the entries of a nonnegative diagonal matrix $D=\textrm{diag}\{ \nu_1,\dots,\nu_n\}$ which, together with a suitable symplectic matrix $M$, permits us to write
\begin{align}
    M^{\top}\left(\frac{\boldsymbol{\sigma}_{(n)}}{\hbar} \right) M&= \left( \begin{array}{ll}
D         &  \mathbf{0}_{n \times n}\\
\mathbf{0}_{n \times n}        & D
    \end{array}\right)\,. \label{symp}
\end{align}
Notice that $\mathcal{S}(\nu_k)=0$ only if $\nu_k=1/2$, and that in the case of one degree of freedom \cite{Holevo1999}, for the particle $a_1$,  we have 
\begin{align}
\nu_1=  \frac{1}{\hbar} \sqrt{   \sigma_{{\bf{p}}_{a_1} {\bf{p}}_{a_1} } \sigma_{{\bf{q}}_{a_1} {\bf{q}}_{a_1} } - \left(\sigma_{{\bf{q}}_{a_1} {\bf{p}}_{a_1} }\right)^2}\,.
\end{align}

It is worth mentioning that using \eqref{symp}, the purity \eqref{eq:qpu} can also be expressed in terms of the symplectic eigenvalues of $\boldsymbol{\sigma}_{(n)} / \hbar$. In fact, we have
\begin{align} \label{pureig}
    \mu \left(a_1,a_2,...,a_n\right)= \left( \frac{1}{2^n}\right)  \prod_{k=1}^n \nu_k^{-1}\,.
\end{align}


\section{Classical analogs of the purity, linear quantum entropy, and von Neumann entropy} \label{subsec:puri}

The goal of this section is to provide classic analogs of the purity \eqref{eq:qpu}, linear quantum entropy \eqref{eq:lS}, and von Neumann entropy \eqref{eq:Sred} in the framework of classical integrable systems. Then, we consider here a subsystem consisting of the $n$ particles $a_1,a_2,...,a_n$ of the classical integrable $N$-system, which is written in terms of the action-angle variables $(\varphi,I)$.

Bearing in mind \eqref{eq:qpu}, the relation \eqref{eq:equicovma} for the quantum $\boldsymbol{\sigma}_{(n)}$ and classical $\sigma_{(n)}^{\mathrm{cl}}$ covariance matrices, i.e., $\boldsymbol{\sigma}_{(n)}\simeq\sigma_{(n)}^{\mathrm{cl}}$, and the Bohr-Sommerfeld quantization rule for the action variables, in the sense $ \hbar/2 \to I_k $, it is natural to define the classical function
\begin{equation}
\mu^{\mathrm{cl}} (a_1,a_2,...,a_n):= \frac{1}{\sqrt{\det \sigma_{(n)}^{\mathrm{cl}}}} \prod^{n}_{k=1} I_{a_k}\,. \label{eq:pca}
\end{equation}
We recall that the $I_{a_k}$ is associated with the $k$-th normal mode. This in turn allows us to define, in analogy with \eqref{eq:lS}, the classical function
\begin{equation}
S^{\textrm{cl}}_L (a_1,a_2,...,a_n):= 1-\mu^{\mathrm{cl}} (a_1,a_2,...,a_n)\,.\label{eq:le}
\end{equation}
It is worth making some comments about \eqref{eq:pca} and \eqref{eq:le}. First, $\mu^{\mathrm{cl}}$ and $S^{\textrm{cl}}_L$ are purely {\it classical} quantities, and  to calculate them, no {\it a priori} knowledge of the corresponding quantum system is required. Second, $\mu^{\mathrm{cl}}$ and $S^{\textrm{cl}}_L$ are functions of the action variables and the system's parameters only. Consequently, to compare $\mu^{\mathrm{cl}}$ and $S^{\textrm{cl}}_L$ with $\mu$ [given by \eqref{eq:qpu}] and $S_L$, respectively, we should resort to the Bohr-Sommerfeld quantization rule and set $I_k=\hbar/2$. However, to avoid the use of this rule and keep the calculations completely in the classical setting, we can define classical functions from \eqref{eq:pca} and \eqref{eq:le} closest to their quantum counterparts by making all action variables equal to a real-positive constant $\alpha$. By doing this, the resulting classical functions are
\begin{subequations}
\begin{align}
\tilde{\mu}^{\mathrm{cl}} (a_1,a_2,...,a_n)&:=\lim_{I_k \to \alpha} \mu^{\mathrm{cl}} (a_1,a_2,...,a_n) \nonumber\\
&=  \alpha^{n}  \lim_{I_k \to \alpha} \frac{1}{\sqrt{\det \sigma_{(n)}^{\mathrm{cl}}}}  \,, \label{eq:pcal}\\
\tilde{S}^{\textrm{cl}}_L(a_1,a_2,...,a_n)&:= 1-\tilde{\mu}^{\mathrm{cl}}(a_1,a_2,...,a_n)\,,\label{eq:leal}
\end{align}
\end{subequations}
which can be regarded as  {\it classical analogs} of the purity \eqref{eq:qpu} and linear quantum entropy \eqref{eq:lS}, respectively. We remark that $\tilde{\mu}^{\mathrm{cl}}$ and $\tilde{S}^{\textrm{cl}}_L$ are classical quantities since in their calculation we do not need to invoke anything from the quantum framework. As we will see in the examples of Sec.~\ref{sec:exam}, the functions $\tilde{\mu}^{\mathrm{cl}}$ and $\tilde{S}^{\textrm{cl}}_L$ yield exactly the same mathematical results as their quantum counterparts and, remarkably, without the need of setting $\alpha= \hbar/2$. Notice that \eqref{eq:pcal} and \eqref{eq:leal} are defined for any integrable system. However, it is only when the corresponding quantum system is in a Gaussian state that $\tilde{\mu}^{\mathrm{cl}}$ and $\tilde{S}^{\textrm{cl}}_L$ correspond to the classical analogs of the quantum purity and linear quantum entropy, respectively. 

In complete analogy with~\eqref{eq:Sred} and taking into account the relation \eqref{eq:equicovma} for the quantum and classical covariance matrices as well as the Bohr-Sommerfeld quantization rule for the action variables $\hbar/2 \to I_k$, we define the classical function
\begin{equation}
    S^{\textrm{cl}} \left( a_1,a_2,...,a_n \right):= \sum_{k=1}^{n} \mathcal{S}^{\textrm{cl}}(\sigma_{k })   \,, \label{Scl}
\end{equation}
where
\begin{align}
    \mathcal{S}^{\textrm{cl}}(\sigma_{k}):=&  \left( \sigma_{k}+\frac{1}{2} \right)\ln \left( \sigma_{k}+\frac{1}{2} \right) \nonumber\\
    &-\left( \sigma_{k}-\frac{1}{2} \right) \ln \left( \sigma_{k}-\frac{1}{2} \right)\,, \label{indientro}
\end{align}
with $\sigma_{k}:= \sigma^{\textrm{cl}}_{k}/2 I_{a_k}$, being $\sigma^{\textrm{cl}}_{k}$ the symplectic eigenvalues of $\sigma^{\textrm{cl}}_{(n)}$. In the case $n=1$, for the particle $a_1$ we get 
\begin{equation}
\sigma_{1}=  \frac{1}{2 I_{a_1}} \sqrt{ \sigma^{\textrm{cl}}_{p_{a_1} p_{a_1} } \sigma^{\textrm{cl}}_{q_{a_1} q_{a_1} } - \left(\sigma^{\textrm{cl}}_{q_{a_1} p_{a_1} }\right)^2}.
\end{equation}

We can go further and define a classical function closer to~\eqref{eq:Sred} by setting all action variables equal to a constant in~\eqref{Scl}. That is, we can define the function 
\begin{subequations}
\begin{align}
   \tilde{S}^{\textrm{cl}} \left( a_1,a_2,...,a_n \right) &:=  \sum_{k=1}^{n} \mathcal{S}^{\textrm{cl}}(\tilde{\sigma}_k) \,, \label{eq:caen}\\
  \tilde{\sigma}_k &:=\lim_{I_k \to \beta} \sigma_k \,, \label{eq:cleige}
\end{align}
\end{subequations}
where $\mathcal{S}^{\textrm{cl}}$ is given by \eqref{indientro} and $\beta$ is a real-positive constant, which, as we will see, disappears during the calculation (as in the classical analog of the purity). The function \eqref{eq:caen} is defined for any integrable system and can be regarded as a  classical analog of the von Neumann entropy when the quantum counterpart of the system under consideration is in a Gaussian state. A remarkable fact about \eqref{eq:caen} is that it can be calculated completely from a classical point of view and, as shown in the examples, it yields exactly the same mathematical results as the von Neumann entropy. 

Notice that the classical analog of the purity \eqref{eq:pca} can also be expressed in terms of the symplectic eigenvalues $\tilde{\sigma}_k$ as
\begin{align}
   \tilde{\mu}^{\mathrm{cl}} (a_1,a_2,...,a_n) = \left( \frac{1}{2^n}\right)  \prod_{k=1}^n \tilde{\sigma}_k^{-1}\,,
\end{align}
in analogy with \eqref{pureig}.

Finally, to better understand why we obtain the same mathematical results of the purity and entropy from a classical point of view, we can observe that we have both local and global information when describing our system in terms of action-angle variables. Thus, using the $n$ action variables, we have a torus $T^n$. In contrast, when analyzing the correlations of the original variables $(q_1(\varphi, I ),\dots, q_n(\varphi, I ),p_1(\varphi, I ) , \dots, p_n(\varphi, I ))$, we see a knot $K^n(K^N)$ since all the variables are correlated in this case and we are observing the $n$-subsystem. In this sense, the classical functions $\tilde{\mu}^{\mathrm{cl}}$, $\tilde{S}^{\textrm{cl}}_L$, and $\tilde{S}^{\textrm{cl}} $ provide a measure of ``non-separability'' of the individual subsystems in phase space.

\section{Examples}\label{sec:exam}

In this section, we present two examples of coupled oscillators for which we compute the classical functions defined in the previous section. At the same time, we compare the results with those found using the quantum definitions for the ground state of the quantum counterpart of these classical integrable systems. Both examples show that our classical approach provides exactly the same results as their quantum counterparts.

\subsection{Three coupled harmonic oscillators}
Let us consider a system of three coupled harmonic oscillators ($N=3$, $a=1,2,3$) with parameters \{$k,k_{12},k_{13}$\}. The Hamiltonian reads
\begin{align}
\begin{split}
H(q,p) =& \frac{1}{2}\left\{ p_1^2 + p_2^2 + p_3^2 + k (q_1^2 + q_2^2 + q_3^2)  \right.\\
& +k_{12} \left[(q_1 - q_2)^2 + (q_2 - q_3)^2\right] \\
&\left.+ k_{13} (q_3 - q_1)^2 \right\}\,.
\end{split}\label{Hejem3pa}
\end{align}
To deal with this Hamiltonian, it is convenient to introduce the transformation from the variables $(q,p)$ to new variables $(Q,P)$ given by
\begin{align} \label{transqQ}
Q&=S q\,,\qquad 
P=S p \,,
\end{align}
where
\begin{align}
Q&= \begin{pmatrix}
 Q_1  \\
 Q_2   \\
  Q_3 
\end{pmatrix} \,,& q&= \begin{pmatrix}
 q_1  \\
 q_2   \\
 q_3 
\end{pmatrix} \,,&
P&= \begin{pmatrix}
 P_1  \\
 P_2   \\
  P_3 
\end{pmatrix} \,,& p&= \begin{pmatrix}
 p_1  \\
 p_2   \\
 p_3 
\end{pmatrix} \,,\end{align}
and
\begin{align}
S=
\begin{pmatrix}
 \frac{1}{\sqrt{3}}   & \frac{1}{\sqrt{3}}  & \frac{1}{\sqrt{3}}\\
 \frac{1}{\sqrt{6}}   & -\sqrt{\frac{2}{3}} & \frac{1}{\sqrt{6}}\\
  -\frac{1}{\sqrt{2}} & 0 & \frac{1}{\sqrt{2}} 
\end{pmatrix} \,,
\end{align}
which allows us to express the Hamiltonian \eqref{Hejem3pa} in the form
\begin{equation}\label{hamiltonianossep}
H (Q,P) = \frac{1}{2}\left(P_1^2 + P_2^2 + P_3^2 + \omega_1^2 Q_1^2 + \omega^2_2 Q_2^2 + \omega_3^2 Q_3^2 \right) \,,
\end{equation}
where
\begin{align}
\omega_1:= \sqrt{k} \,, \,\, \omega_2:= \sqrt{k + 3 k_{12}} \,, \,\, \omega_3:= \sqrt{k + k_{12} + 2 k_{13}} \,,\label{frec}
\end{align}
are the frequencies of uncoupled harmonic oscillators. Notice that we restricted ourselves to the case $k>0$, $k + 3 k_{12}>0,$ and $k + k_{12} + 2 k_{31}>0$. In turn, the transformation from the variables $(Q,P)$ to the action-angle variables $(\varphi,I)$ is given by
\begin{equation}\label{sco:Q0P0}
Q_{a}=\left(\frac{2 I_a}{\omega_a}\right)^{1/2} \! \sin\varphi_{a}\,,  \ \ P_{a}=\left(2 \omega_a I_a\right)^{1/2} \cos\varphi_{a}\, ,
\end{equation}
and the classical average of a function $f=f(\varphi,I)$ is computed via
\begin{equation}
\langle f \rangle_{\mathrm{cl}}= \frac{1}{\left( 2 \pi\right)^3}   \iiint_{0}^{2 \pi}  \!\!\! \mathrm{d}^3 \varphi \, f  \,.
\end{equation}

With this at hand, we compute the $6\times 6$ classical covariance matrix \eqref{eq:sigmac}, obtaining
\begin{align}\label{qqppcla}
\sigma^{\textrm{cl}}&= \left(
\begin{array}{cc}
\sigma_{qq} & \sigma_{qp} \\
 \sigma_{qp} & \sigma_{pp}
\end{array}
\right)\,,
\end{align}
where the block matrices $\sigma_{qq}$, $\sigma_{pp}$, and $\sigma_{qp}$ are given by
\begin{widetext}
\begin{subequations}  \allowdisplaybreaks
\begin{align}
\sigma_{qq}&= \frac{1}{3}\left(
\begin{array}{ccc}
 \frac{1}{2} \left(\frac{2 I_1}{\omega_1}+\frac{I_2}{\omega_2}+\frac{3 I_3}{\omega_3}\right) & \frac{I_1}{\omega_1}-\frac{I_2}{\omega_2} & \frac{1}{2} \left(\frac{2 I_1}{\omega_1}+\frac{I_2}{\omega_2}-\frac{3 I_3}{\omega_3}\right) \\
\frac{I_1}{\omega_1}-\frac{I_2}{\omega_2} & \frac{I_1}{\omega_1}+\frac{2 I_2}{\omega_2} & \frac{I_1}{\omega_1}-\frac{I_2}{\omega_2} \\
 \frac{1}{2} \left(\frac{2 I_1}{\omega_1}+\frac{I_2}{\omega_2}-\frac{3 I_3}{\omega_3}\right) & \frac{I_1}{\omega_1}-\frac{I_2}{\omega_2} & \frac{1}{2} \left(\frac{2 I_1}{\omega_1}+\frac{I_2}{\omega_2}+\frac{3 I_3}{\omega_3}\right) \\
\end{array}
\right) \,,\\
\sigma_{pp}&=  \frac{1}{3}
\left(
\begin{array}{ccc}
 \frac{1}{2} (2 I_1 \omega_1+I_2 \omega_2+3 I_3 \omega_3) & I_1 \omega_1-I_2 \omega_2 & \frac{1}{2} (2 I_1 \omega_1+I_2 \omega_2-3 I_3 \omega_3) \\
 I_1 \omega_1-I_2 \omega_2 & I_1 \omega_1+2 I_2 \omega_2 & I_1 \omega_1-I_2 \omega_2 \\
 \frac{1}{2} (2 I_1 \omega_1+I_2 \omega_2-3 I_3 \omega_3) & I_1 \omega_1-I_2 \omega_2 & \frac{1}{2} (2 I_1 \omega_1+I_2 \omega_2+3 I_3 \omega_3) \\
\end{array}
\right)\,,\\
\sigma_{qp}&= 
\mathbf{0}_{3 \times 3}\,.
\end{align}
\end{subequations}
We can obtain the classical covariance matrix of each subsystem by taking from \eqref{qqppcla} the respective rows and columns. The subsystems are as follows: each oscillator $(1)$, $(2),$ and $(3)$; the possible pairs $(1,2)$, $(2,3),$ and $(1,3)$; and the complete system $(1,2,3)$. 

\subsubsection{Classical analog of purity}

Using \eqref{eq:pca} and \eqref{qqppcla}, we obtain the following results for each subsystem:
\begin{subequations}  \allowdisplaybreaks\label{purezascla}
\begin{align}
\mu^{\mathrm{cl}}(1)&= 6 I_1 \sqrt{\frac{\omega_1 \omega_2 \omega_3}{(2 I_1\omega_1+I_2\omega_2+3 I_3\omega_3) \left[\omega_1 (3 I_3 \omega_2+I_2\omega_3)+2 I_1 \omega_2 \omega_3\right]}} \,,\\
\mu^{\mathrm{cl}}(2)&= 3 I_2 \sqrt{\frac{\omega_1 \omega_2}{(I_1\omega_2+2I_2 \omega_1) (I_1\omega_1+2 I_2 \omega_2)}} \,,\\
\mu^{\mathrm{cl}}(3)&=  
\frac{I_3}{I_1} \mu^{\mathrm{cl}}(1)\,,\\
\mu^{\mathrm{cl}}(1,2)&= 6 I_1 I_2\sqrt{\frac{\omega_1 \omega_2 \omega_3}{(2 I_2 I_3 \omega_1+ I_1 I_3\omega_2+3 I_1 I_2 \omega_3) \left[I_1 \omega_1 (3 I_2\omega_2+I_3\omega_3)+2 I_2 I_3\omega_2 \omega_3\right]}} \,,\\
  \mu^{\mathrm{cl}}(1,3)&=   3 I_1 \sqrt{\frac{\omega_1 \omega_2}{(2 I_1 \omega_2+ I_2\omega_1) (2 I_1\omega_1+I_2\omega_2)}}\,,\\
  \mu^{\mathrm{cl}}(2,3)&=  \frac{I_3}{I_1}  \mu^{\mathrm{cl}}(1,2) \,,\\
  \mu^{\mathrm{cl}}(1,2,3)&=1\,.\label{purezasclafull}
\end{align}
\end{subequations}
\end{widetext}
Notice that the resulting functions $\mu^{\mathrm{cl}}$ depend on the action variables $I$ and the normal frequencies $\omega$, except in the case of the complete system [see \eqref{purezasclafull}].

We now calculate the classical analog of the purity \eqref{eq:pcal}. Setting $I_1=I_2=I_3=\alpha$ in \eqref{purezascla}, it is straightforward to obtain
\begin{subequations} \label{purezasclacla}
\begin{align}
\tilde{\mu}^{\mathrm{cl}}(1)&= 6  \sqrt{\frac{\omega_1 \omega_2 \omega_3}{(2 \omega_1+\omega_2+3 \omega_3) \left[\omega_1 (3  \omega_2+\omega_3)+2 \omega_2 \omega_3\right]}} \,,\\
\tilde{\mu}^{\mathrm{cl}}(2)&= 3  \sqrt{\frac{\omega_1 \omega_2}{(\omega_2+2 \omega_1) (\omega_1+2 \omega_2)}} \,,
\end{align}
\end{subequations}
along with $\tilde{\mu}^{\mathrm{cl}}(3)= \tilde{\mu}^{\mathrm{cl}}(1,2)= \tilde{\mu}^{\mathrm{cl}}(2,3)=\tilde{\mu}^{\mathrm{cl}}(1)$, $\tilde{\mu}^{\mathrm{cl}}(1,3)=\tilde{\mu}^{\mathrm{cl}}(2)$, and $\tilde{\mu}^{\mathrm{cl}}(1,2,3)=\mu^{\mathrm{cl}}(1,2,3)$. Some of these equalities are expected because the system is symmetric under the interchange $1\leftrightarrow 3$. Let us make some comments about these results.  
\begin{itemize}
    \item As expected the functions $\tilde{\mu}^{\mathrm{cl}}$ take values between $0$ and $1$. Notice that they do not depend on $\alpha$.
    \item Notice that if $\omega_1=\omega_2=\omega_3$, then $\tilde{\mu}^{\mathrm{cl}}(1)=1$, which is the case for uncoupled oscillators $k_{12}=k_{13}=0$. This means that the oscillator ($1$) is pure only if it does not interact with the oscillators (2) and (3), as we can expect.
    \item We have that  if $\omega_1=\omega_2$, then $\tilde{\mu}^{\mathrm{cl}}(2)=1$, which corresponds to $k_{12}=0$. This means that the oscillator ($2$) is pure if it does not interact with the oscillators (1) and (3).
    \item Using these classical analogs of purity and \eqref{eq:leal} we can also calculate the classical analog of the linear quantum entropy.
\end{itemize}
\vspace{.5cm}

\subsubsection{Classical analog of the von Neumann entropy}
\begin{figure}[h!]
\includegraphics[width=.4\textwidth]{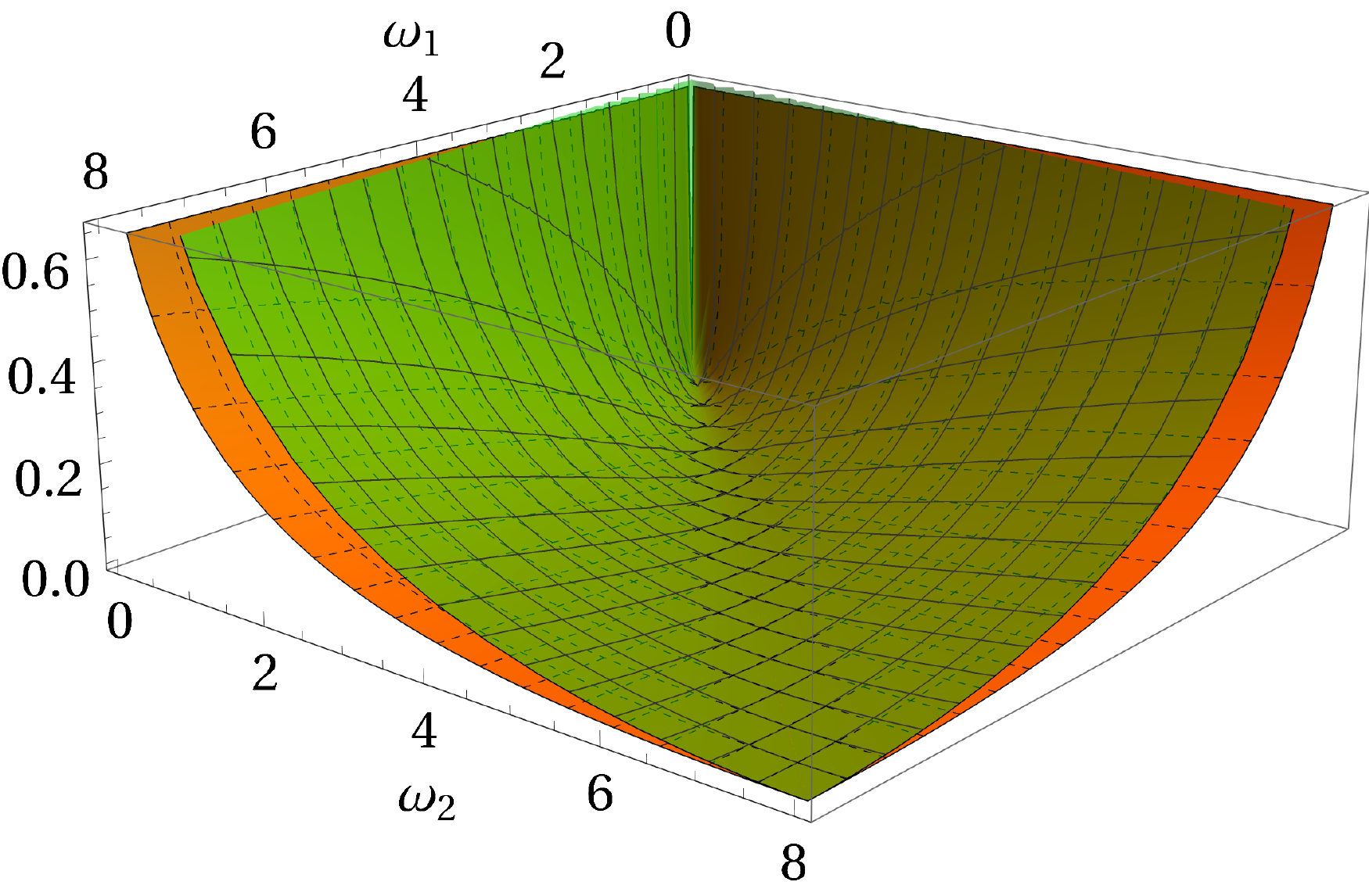} \\
(a) \\
\includegraphics[width=.4\textwidth]{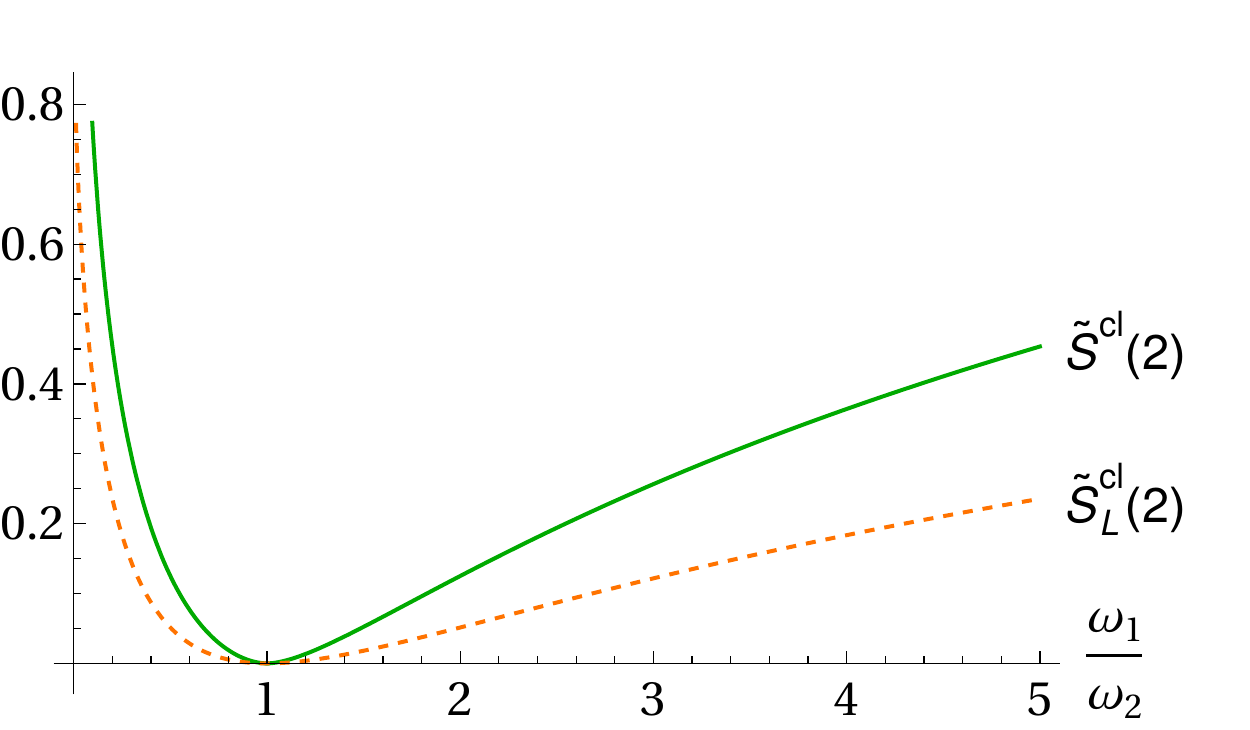} \\
(b)
\caption{(a) Classical analog of the quantum entropy $\tilde{S}^{\mathrm{cl}}(2)$ (green) and classical analog of the linear quantum entropy $\tilde{S}_L^{\mathrm{cl}}(2)$ (dashed orange) for the subsystem (2) as functions of $\omega_1$ and $\omega_2$. (b) $\tilde{S}^{\mathrm{cl}}(2)$ (solid green) and $\tilde{S}_L^{\mathrm{cl}}(2)$ (dashed orange), but now as functions of $\omega_1/\omega_2$.}
\label{Figures}
\end{figure}

To calculate the classical analog of the entropy we need the symplectic eigenvalues $\sigma_{a_k}$ of the classical covariance matrix of each subsystem and the corresponding $\tilde{\sigma}_{a_k}$. They are given by
\begin{widetext}
\begin{subequations} \allowdisplaybreaks\label{eq:eigexa}
\begin{align}
(1):&  \, \sigma_1=  \frac{1}{12 I_1} \sqrt{\left(\frac{2 I_1}{\omega_1}+\frac{I_2}{\omega_2}+\frac{3 I_3}{\omega_3}\right) (2 I_1 \omega_1+I_2 \omega_2+3 I_3 \omega_3)}  \nonumber\\
&\Rightarrow \tilde{\sigma}_1= \frac{1}{12} \sqrt{\left(\frac{2 }{\omega_1}+\frac{1}{\omega_2}+\frac{3}{\omega_3}\right) (2 \omega_1+ \omega_2+3 \omega_3)}\,, \\
(2):&  \, \sigma_1= \frac{1}{6 I_2} \sqrt{\left(\frac{I_1}{\omega_1}+\frac{2 I_2}{\omega_2}\right) (I_1 \omega_1+2 I_2 \omega_2)}   \Rightarrow \tilde{\sigma}_1=  \frac{1}{6} \sqrt{5+\frac{2 \omega_1}{\omega_2}+\frac{2 \omega_2}{\omega_1}}\,, \\
(3):&  \, \sigma_1=  \frac{1}{12 I_3} \sqrt{\left(\frac{2 I_1}{\omega_1}+\frac{I_2}{\omega_2}+\frac{3 I_3}{\omega_3}\right) (2 I_1 \omega_1+I_2 \omega_2+3 I_3 \omega_3)} \nonumber \\
&\Rightarrow \tilde{\sigma}_1= \frac{1}{12} \sqrt{\left(\frac{2 }{\omega_1}+\frac{1}{\omega_2}+\frac{3}{\omega_3}\right) (2 \omega_1+ \omega_2+3 \omega_3)}\,, \\
(1,2):& \left\{ \begin{array}{c}
   \sigma_1 =  \frac{1}{12 I_1} \sqrt{ \frac{1}{2 \omega_1 \omega_2 \omega_3}\left( E- \sqrt{F+E^2}\right)} \\
    \sigma_2 =   \frac{1}{12 I_2} \sqrt{ \frac{1}{2\omega_1 \omega_2 \omega_3}\left( E+ \sqrt{F+E^2}\right)} \\
\end{array}\right. \nonumber \\
&\Rightarrow \begin{array}{l}
   \tilde{\sigma}_1  =  \frac{1}{2}\\
    \tilde{\sigma}_2 = \frac{1}{12} \sqrt{\left(\frac{2 }{\omega_1}+\frac{1}{\omega_2}+\frac{3}{\omega_3}\right) (2 \omega_1+ \omega_2+3 \omega_3)}
\end{array} \,,\\
\textrm{with } & E:= 16 I_1^2 \omega_1 \omega_2 \omega_3+2 I_1 I_2 \omega_3 \left(\omega_1^2+\omega_2^2\right)+6 I_1 I_3 \omega_2 \left(\omega_1^2+\omega_3^2\right)+25 I_2^2 \omega_1 \omega_2 \omega_3 \nonumber\\
 & \, \quad +3 I_2 I_3 \omega_1 \left(\omega_2^2+\omega_3^2\right)+9 I_3^2 \omega_1 \omega_2 \omega_3  \,, \nonumber\\
 & F:= -144 \omega_1 \omega_2 \omega_3 (3 I_1 I_2 \omega_3+I_1 I_3 \omega_2+2 I_2 I_3 \omega_1) \left[I_1 \omega_1 (3 I_2 \omega_2+I_3 \omega_3)+2 I_2 I_3 \omega_2 \omega_3\right] \,, \nonumber\\
(1,3):& \left\{ \begin{array}{l}
   \sigma_1  =  \frac{I_3}{2I_1} \\
    \sigma_2 = \frac{1}{6 I_3} \sqrt{\left( \frac{2 I_1}{\omega_1} +\frac{I_2}{ \omega_2}\right) (2 I_1 \omega_1+I_2 \omega_2)}
\end{array}\right. \Rightarrow \begin{array}{l}
   \tilde{\sigma}_1  =  \frac{1}{2}\\
    \tilde{\sigma}_2 =   \frac{1}{6} \sqrt{5+\frac{2 \omega_1}{\omega_2}+\frac{2 \omega_2}{\omega_1}}
\end{array}  \,,\\
(2,3):& \left\{ \begin{array}{l}
   \sigma_1  =   \frac{1}{12 I_2} \sqrt{ \frac{1}{2 \omega_1 \omega_2 \omega_3}\left( E- \sqrt{F+E^2}\right)} \\
    \sigma_2 = \frac{1}{12 I_3} \sqrt{ \frac{1}{2 \omega_1 \omega_2 \omega_3}\left( E+ \sqrt{F+E^2}\right)}
\end{array}\right. \nonumber\\
&\Rightarrow \begin{array}{l}
   \tilde{\sigma}_1  = \frac{1}{2} \\
    \tilde{\sigma}_2 = \frac{1}{12} \sqrt{\left(\frac{2 }{\omega_1}+\frac{1}{\omega_2}+\frac{3}{\omega_3}\right) (2 \omega_1+ \omega_2+3 \omega_3)}
\end{array} \,,\\
(1,2,3):& \left\{ \begin{array}{c}
   \sigma_1 = \frac{1}{2}=\tilde{\sigma}_1 \\
    \sigma_2 =\frac{1}{2}= \tilde{\sigma}_2 \\
    \sigma_3 =\frac{1}{2}= \tilde{\sigma}_3
\end{array}\right. \,.
\end{align}
\end{subequations}

Using these results, the classical analogs of the von Neumann entropy \eqref{eq:caen} for each subsystem are
\begin{subequations} \allowdisplaybreaks \label{ex1-class-entro}
\begin{align}
   \tilde{S}^{\mathrm{cl}}(1)& = \frac{1}{12} \left(\log \left\{\frac{\left[2 \omega_3 \left(\omega_1^2-11 \omega_1 \omega_2+\omega_2^2\right)+3 \omega_3^2 (\omega_1+2 \omega_2)+3 \omega_1 \omega_2 (2 \omega_1+\omega_2)\right]^6}{8916100448256 \omega_1^6 \omega_2^6 \omega_3^6}\right\} \right.\nonumber\\
   &\left. \!\!\!\!\!\!\!\!\!\!\!\!\!\!\!+2 \sqrt{\left(\frac{2}{\omega_1}+\frac{1}{\omega_2}+\frac{3}{\omega_3}\right) (2 \omega_1+\omega_2+3 \omega_3)} \tanh ^{-1}\left(\frac{6}{\sqrt{\left(\frac{2}{\omega_1}+\frac{1}{\omega_2}+\frac{3}{\omega_3}\right) (2 \omega_1+\omega_2+3 \omega_3)}}\right)\right) \,,\\
   \tilde{S}^{\mathrm{cl}}(2)&=\frac{1}{6} \left(2 \sqrt{\frac{2 \omega_1}{\omega_2}+\frac{2 \omega_2}{\omega_1}+5} \tanh ^{-1}\left(\frac{3}{\sqrt{\frac{2 \omega_1}{\omega_2}+\frac{2 \omega_2}{\omega_1}+5}}\right)-3 \log \left(\frac{18 \omega_1 \omega_2}{(\omega_1-\omega_2)^2}\right)\right)\,,
\end{align}
\end{subequations}
\end{widetext}
along with $\tilde{S}^{\mathrm{cl}}(3)= \tilde{S}^{\mathrm{cl}}(1,2)= \tilde{S}^{\mathrm{cl}}(2,3)=\tilde{S}^{\mathrm{cl}}(1)$, $\tilde{S}^{\mathrm{cl}}(1,3)= \tilde{S}^{\mathrm{cl}}(2)$, and $\tilde{S}^{\mathrm{cl}}(1,2,3)=0$. Notice that if $\omega_1=\omega_2=\omega_3$ then $\tilde{S}^{\mathrm{cl}}(1)=0$, and if $\omega_1=\omega_2$ then $\tilde{S}^{\mathrm{cl}}(2)=0$. To illustrate the behavior of these functions, in Fig. \ref{Figures} (a) we plot the classical analog of the entropy $\tilde{S}^{\mathrm{cl}}(2)$ and the classical analog of linear quantum entropy $\tilde{S}_L^{\mathrm{cl}}(2)$ as functions of the frequencies $\omega_1$ and $\omega_2$. Also, noticing that $\tilde{S}^{\mathrm{cl}}(2)$ and $\tilde{S}_L^{\mathrm{cl}}(2)$ actually depend on the quotient $\omega_1/\omega_2$, in Fig. \ref{Figures} (b) we plot them as functions of this quotient. Clearly, we see that both $\tilde{S}^{\mathrm{cl}}(2)$ and $\tilde{S}_L^{\mathrm{cl}}(2)$ are equal to zero when $\omega_1=\omega_2$, as we already pointed out. 

\subsubsection{Quantum approach}
The aim of this subsection is to compare the previous results with those coming from the quantum approach. The quantum counterpart of the Hamiltonian \eqref{Hejem3pa} is
\begin{align}
\begin{split}
\hat{\bf{H}}(\hat{\bf{q}},\hat{\bf{p}}) &= \frac{1}{2}\left\{ \hat{\bf{p}}_1^2 + \hat{\bf{p}}_2^2 + \hat{\bf{p}}_3^2 + k (\hat{\bf{q}}_1^2 + \hat{\bf{q}}_2^2 + \hat{\bf{q}}_3^2)  \right.\\
&\qquad  +k_{12} \left[(\hat{\bf{q}}_1 - \hat{\bf{q}}_2)^2 + (\hat{\bf{q}}_2 - \hat{\bf{q}}_3)^2\right] \\
&\left.\qquad+ k_{13} (\hat{\bf{q}}_3 - \hat{\bf{q}}_1)^2 \right\}\,.
\end{split}
\end{align}
The ground-state wave function of this system is Gaussian and given by
\begin{align}
\begin{split}
\psi_0(q)=& \frac{(\omega_1 \omega_2 \omega_3)^{1/4}}{\pi^{3/4} \hbar ^{3/4}}  \exp \left\{-\frac{1}{12 \hbar } \left[ 2 q_2^2 (\omega_1+2 \omega_2)  \right. \right.\\
& +(q_1^2+q_3^2) (2 \omega_1+\omega_2+3 \omega_3)\\
&+4 q_2 (q_1 +q_3)(\omega_1-\omega_2) \\
&\left. \left.+2 q_1 q_3 (2 \omega_1+\omega_2-3 \omega_3) \right]\right\}\,.
\end{split}
\end{align}
Using this state, the quantum covariance matrix \eqref{qumet} turns out to be
\begin{align}\label{covarianceq}
\boldsymbol{\sigma}&= \left(
\begin{array}{cc}
\boldsymbol{\sigma}_{qq} & \boldsymbol{\sigma}_{qp} \\
 \boldsymbol{\sigma}_{qp} & \boldsymbol{\sigma}_{pp}
\end{array}
\right)\,
\end{align}
where
\begin{widetext}
\begin{subequations} \allowdisplaybreaks \label{covarianceq2}
\begin{align}
\boldsymbol{\sigma}_{qq}&= \frac{\hbar}{6} \left(
\begin{array}{ccc}
 \frac{1}{2} \left(\frac{2}{\omega_1}+\frac{1}{\omega_2}+\frac{3 }{\omega_3}\right) & \frac{1}{\omega_1}-\frac{1}{\omega_2} & \frac{1}{2} \left(\frac{2 }{\omega_1}+\frac{1}{\omega_2}-\frac{3 }{\omega_3}\right) \\
 \frac{1}{\omega_1}-\frac{1}{\omega_2} & \frac{1}{\omega_1}+\frac{2 }{\omega_2} & \frac{1}{\omega_1}-\frac{1}{\omega_2} \\
 \frac{1}{2} \left(\frac{2 }{\omega_1}+\frac{1}{\omega_2}-\frac{3 }{\omega_3}\right) & \frac{1}{\omega_1}-\frac{1}{\omega_2} & \frac{1}{2} \left(\frac{2 }{\omega_1}+\frac{1}{\omega_2}+\frac{3}{\omega_3}\right) \\
\end{array}
\right)\,,\\
\boldsymbol{\sigma}_{pp}&= \frac{\hbar}{6}
\left(
\begin{array}{ccc}
 \frac{1}{2} (2 \omega_1+ \omega_2+3  \omega_3) &  \omega_1- \omega_2 & \frac{1}{2} (2 \omega_1+ \omega_2-3 \omega_3) \\
  \omega_1- \omega_2 &  \omega_1+2  \omega_2 &  \omega_1- \omega_2 \\
 \frac{1}{2} (2  \omega_1+ \omega_2-3  \omega_3) &  \omega_1- \omega_2 & \frac{1}{2} (2  \omega_1+ \omega_2+3 \omega_3) \\
\end{array}
\right)\,,\\
\boldsymbol{\sigma}_{qp}&= \mathbf{0}_{3\times 3}\,.
\end{align}
\end{subequations}

By using the Bohr-Sommerfeld quantization rule $I_a \to \hbar/2$, it is straightforward to verify that the resulting classical and quantum covariant matrices, \eqref{qqppcla} and \eqref{covarianceq}, satisfy the relation \eqref{eq:equicovma}.

The covariance matrix of each subsystem can be easily obtained by taking from the above matrix the respective rows and columns. Using \eqref{eq:qpu} and \eqref{covarianceq}, we compute the purity of each subsystem, obtaining
\begin{subequations}  \allowdisplaybreaks \label{purezas}
\begin{align}
\mu(1)&= 6\, \sqrt{\frac{\omega_1 \omega_2 \omega_3}{(2 \omega_1+\omega_2+3 \omega_3) [\omega_1 (3 \omega_2+\omega_3)+2 \omega_2 \omega_3]}} \,,\\
\mu(2)&= 3 \, \sqrt{\frac{\omega_1 \omega_2}{(2 \omega_1+\omega_2) (\omega_1+2 \omega_2)}} \,,
\end{align}
\end{subequations}
\end{widetext}
along with $\mu(1,2,3)=1$, $\mu(3)= \mu(1,2)= \mu(2,3)=\mu(1)$, and $\mu(1,3)=\mu(2)$. Comparing \eqref{purezasclacla} and \eqref{purezas}, it is direct to see that both results are exactly the same. Hence, in this example, we verified that the classical function $\tilde{\mu}^{\mathrm{cl}}$ is able to produce the same mathematical results of the quantum purity $\mu$.  

We now turn to the calculation of the von Neumann entropy. Using \eqref{covarianceq}, it can be verified that the symplectic eigenvalues $\nu_k$ of the corresponding matrices $\boldsymbol{\sigma}_{(n)} / \hbar$ are equal to the eigenvalues $\tilde{\sigma}_k$ given by \eqref{eq:eigexa}. This means that, from \eqref{eq:Sred}, the resulting von Neumann entropies of the subsystems are the same as determined from the classical function \eqref{eq:caen} with \eqref{eq:eigexa}, i.e., they are given by \eqref{ex1-class-entro}. This corroborates that the classical function \eqref{eq:caen} is capable of giving rise to the same mathematical results of the von Neumann entropy.

It is worthwhile mentioning that in~\cite{Simon2000} (see \cite{Duan2000} for an equivalent criterion), Simon introduced a separability criterion for Gaussian states of a bipartite system of two harmonic oscillators. Soon after, in \cite{Werner2001} (see also \cite{Giedke2001, Simon2003}), a separability criterion for Gaussian states of $1\times M$ ($M$ is arbitrary large) oscillators was provided, which is the case of this example. As explained in these references, the Gaussian states are entangled if the criteria are violated. These criteria are linked to our results by the observation (condition) that only when some of the normal frequencies are equal, the corresponding subsystems are pure and the von Neumann entropy is zero. Certainly, in that cases, the subsystems are separable and classically uncoupled.

Before concluding this example it is important to emphasize that our classical analogs of the purity, linear quantum entropy, and von Neumann entropy are completely defined in the classical context, i.e., we do not need any {\it a priori} information from the quantum context to compute them.

\subsection{Linearly coupled harmonic oscillators}

We now want to illustrate our classical approach on a system of two coupled harmonic oscillators described by the Hamiltonian
\begin{equation}
H(q,p)= \frac{1}{2} \left( p^2_1+p^2_2 + A q_1^2+B q_2^2+ C q_1 q_2 \right)\,, \label{H2ch}
\end{equation}
where $A,B >0$ are parameters and we restrict our study to the region $A \neq B,  4AB-C\geq 0$. This system has been widely analyzed  in the context of quantum entanglement and, in particular, it has been shown that for certain  parameter values it exhibits a very large quantum entanglement \cite{Jaeger2007, Makarov2018, Han99}. Also, in \cite{Alvarez2019}  this model was used to study the classical analog of the quantum geometric tensor \eqref{eq:defQGT} restricted to the parameter space. Interestingly, it was shown that the classical analog of the quantum metric tensor does not yield the full parameter structure of its quantum counterpart, the cause being a quantum anomaly. The origin of such anomaly was that the transformation matrix, which brings the system into its normal form, depends on the system's parameters \cite{Gonzalez2020}. In contrast, we show in this subsection that our classical analogs exactly reproduce the mathematical results of their quantum counterparts.

To bring the system \eqref{H2ch} into its normal form, we perform the canonical transformation 
\begin{align} \label{transqQ22}
Q&=R q\,,\qquad 
P=R p \,,
\end{align}
where
\begin{align}
Q&= \begin{pmatrix}
 Q_1  \\
 Q_2   
\end{pmatrix} \,,& q&= \begin{pmatrix}
 q_1  \\
 q_2  
\end{pmatrix} \,,&
P&= \begin{pmatrix}
 P_1  \\
 P_2   
\end{pmatrix} \,,& p&= \begin{pmatrix}
 p_1  \\
 p_2  
\end{pmatrix} \,,\end{align}
and
\begin{align}
R&=  \left(
\begin{array}{cc}
 \cos \alpha &-\sin \alpha \\
 \sin \alpha & \cos \alpha \\
\end{array}
\right)\,.
\end{align}
Here, the angle $\alpha$ is such that $\tan 2\alpha = C/(B-A)$ with $\alpha\in (-\pi/4,\pi/4)$. Notice that the  transformation matrix $R$ depends on the system's parameters. In terms of the new variables, the Hamiltonian reads
\begin{equation}\label{hamiltonianossep22}
H (Q,P) = \frac{1}{2}\left(P_1^2 + P_2^2 + \omega_1^2 Q_1^2 + \omega^2_2 Q_2^2 \right) \,,
\end{equation}
where $\omega_1$ and $\omega_2$ are the normal frequencies
\begin{align}
    \omega_1:=\sqrt{A-\frac{C}{2} \tan \alpha} \,, \qquad    \omega_2:=\sqrt{B+\frac{C}{2} \tan \alpha} \,.
\end{align}
We further express the new coordinates in terms of the action-angle variables $(\varphi_{a},I_a)$ as
\begin{equation}\label{eq:qpith}
Q_{a}=\left(\frac{2 I_a}{\omega_a}\right)^{1/2}\sin\varphi_{a}\,,  \ \ P_{a}=\left(2 \omega_a I_a\right)^{1/2} \cos\varphi_{a}\, ,
\end{equation}
and compute the classical average via
\begin{equation}
\langle f \rangle_{\mathrm{cl}}= \frac{1}{\left( 2 \pi\right)^2}    \iint_{0}^{2 \pi} \!\!\! \mathrm{d}^2 \varphi \, f  \,.
\end{equation}

Using \eqref{eq:sigmac} and the previous transformations, we arrive at the $4\times 4$ classical covariance matrix $\sigma^{\textrm{cl}}$ of the system:
\begin{align}\label{qqppcla22}
    \sigma^{\textrm{cl}}&= \left(
\begin{array}{cc}
\sigma_{qq} & \sigma_{qp} \\
 \sigma_{qp} & \sigma_{pp}
\end{array}
\right)\,,
\end{align}
where
\begin{widetext}
\begin{subequations}\allowdisplaybreaks
\begin{align}
 \sigma_{q q} &=  \left(
\begin{array}{cc}
 \frac{I_1 \cos^2 \alpha}{\omega_1}+\frac{I_2 \sin^2 \alpha}{\omega_2} &  \left( \frac{I_2}{\omega_2}-\frac{I_1}{\omega_1} \right)\sin \alpha \cos \alpha \\
 \left( \frac{I_2}{\omega_2}-\frac{I_1}{\omega_1} \right)\sin \alpha \cos \alpha  &  \frac{I_1 \sin^2 \alpha}{\omega_1}+\frac{I_2\cos^2 \alpha}{\omega_2} \\
\end{array}
\right)\,,\\
 \sigma_{ p p}  &=  \left(
\begin{array}{cc}
 I_1 \omega_1 \cos^2 \alpha+ I_2\omega_2 \sin^2 \alpha        &  \left( I_2\omega_2-I_1\omega_1\right) \sin \alpha \cos \alpha \\
\left( I_2\omega_2-I_1\omega_1\right) \sin \alpha \cos \alpha &  I_1\omega_1 \sin^2 \alpha+ I_2\omega_2 \cos^2 \alpha \\
\end{array}
\right)\,,\\
\sigma_{qp}&= 
\mathbf{0}_{2 \times 2}\,.
\end{align}
\end{subequations}
In this case, the subsystems are: each oscillator (1) and (2); and the complete system (1,2). With \eqref{qqppcla22} at hand, we compute the classical analog of the purity \eqref{eq:pcal}. The result for each subsystem is
\begin{subequations}\label{purity-maka}
\begin{align}
\tilde{\mu}^{\mathrm{cl}}(1)&= \sqrt{ \frac{ \omega_1 \omega_2}{\left( \omega_1\sin^2 \alpha+\omega_2\cos^2 \alpha\right) \left(\omega_1 \cos^2 \alpha+\omega_2 \sin^2 \alpha \right)}} =  \sqrt{  \frac{4 A B-C^2}{4A B}} =  \tilde{\mu}^{\mathrm{cl}}(2) \,,\\
\tilde{\mu}^{\mathrm{cl}}(1,2)&=  1\,.
\end{align}
\end{subequations}
Notice that if $C=0$ then $\tilde{\mu}^{\mathrm{cl}}(1)=1=\tilde{\mu}^{\mathrm{cl}}(2)$, which means that the oscillators (1) and (2) are pure only if they are uncoupled, as expected. Remarkably, the resulting classical purities \eqref{purity-maka} exactly match their quantum counterparts, which were calculated in \cite{Han99} for the ground-state wave function of the corresponding quantum system.

On the other hand, the symplectic eigenvalue of the subsystem $(1)$ is
\begin{align}
(1): \qquad  \sigma_1&=\frac{1}{2 I_1} \sqrt{ \left( \frac{I_1\cos^2 \alpha }{\omega_1}+\frac{I_2\sin ^2 \alpha}{\omega_2}\right) \left(I_1 \omega_1 \cos^2 \alpha+ I_2 \omega_2 \sin^2 \alpha \right)} \nonumber\\
\Rightarrow
\tilde{\sigma}_1 &=\frac{1}{2} \sqrt{\left(\frac{\cos^2 \alpha }{\omega_1}+\frac{\sin ^2 \alpha}{\omega_2}\right) \left(\omega_1 \cos^2 \alpha+ \omega_2 \sin^2 \alpha \right)} = \sqrt{\frac{A B}{4A B- C^2}} \,, \label{eH2}
\end{align}
and, because of the symmetry between the subsystems $(1)$ and $(2)$, the symplectic eigenvalue of the subsystem $(2)$ is $\tilde{\sigma}_2=\tilde{\sigma}_1$. Furthermore, the symplectic eigenvalues of the complete system $(1,2)$ are $ \sigma_1=1/2= \sigma_2=\tilde{\sigma}_1=\tilde{\sigma}_2$. Then, the classical analogs of the von Neumann entropy for the subsystems $(1)$, $(2)$, and $(1,2)$ turn out to be
\begin{subequations}\label{entropy-maka}
\begin{align}
\tilde{S}^{\mathrm{cl}}(1) =& \left(  \sqrt{\frac{A B}{4A B- C^2}}+\frac{1}{2} \right)\ln \left(  \sqrt{\frac{A B}{4A B- C^2}}+\frac{1}{2} \right)  
-\left(  \sqrt{\frac{A B}{4A B- C^2}}-\frac{1}{2} \right) \ln \left(  \sqrt{\frac{A B}{4A B- C^2}}-\frac{1}{2} \right) =\tilde{S}^{\mathrm{cl}}(2)\,,\\
\tilde{S}^{\mathrm{cl}}(1,2)=&0 \,.
\end{align}
\end{subequations}
\end{widetext}
Notice that if $C=0$ then $\tilde{S}^{\mathrm{cl}}(1)=\tilde{S}^{\mathrm{cl}}(2)=0$, i.e., they vanish when the oscillators are uncoupled. Finally, it can be verified that the resulting classical entropies  \eqref{entropy-maka} exactly match their quantum counterparts for the ground state of the quantum version of \eqref{H2ch}.

\section{Conclusions}\label{sec:conclu}

In this paper, we have first derived the classical covariance matrix \eqref{eq:sigmac} for integrable systems, which is the classical analog of the quantum covariance matrix, as is stated by the relation \eqref{eq:equicovma}. We have achieved this classical covariance matrix by using the classical limit of the Wigner function and shown that the relation \eqref{eq:equicovma} between both covariance matrices holds even if the corresponding quantum state is not Gaussian, which was illustrated in the case of the anharmonic oscillator. Then, we have introduced the functions \eqref{eq:pcal}, \eqref{eq:leal}, and \eqref{eq:caen} for classical integrable systems and shown through examples that they are, respectively, the classical analogs of the purity, linear quantum entropy, and von Neumann entropy as long as the quantum counterpart of the system is in a Gaussian state. The classical functions \eqref{eq:pcal}, \eqref{eq:leal}, and \eqref{eq:caen} provide, respectively, a mathematical description of the purity, linear quantum entropy, and von Neumann entropy of the system from a classical point of view, without resorting to quantum context. These classical functions depend on the classical covariance matrix which only involves averages over the angle variables of the system. Our classical analogs reveal how much information from the complete system remains in the subsystem under consideration. As we have explained, this is because by using the action-angle variables we have both local and global information.

To illustrate the approach, we have calculated our classical analogs for a nontrivial system of three coupled harmonic oscillators, and for a system of two linear coupled oscillators which has been previously analyzed in the literature. We have found in these examples that the aforementioned classical analogs exactly matched their quantum versions.

Let us end with some comments. (i) Our classical analogs may help to provide more insight into the investigation of separability of quantum states. In this line of thought, our approach may be exploited to develop a classical analog of the separability criterion for Gaussian states  \cite{Simon2000,Duan2000,Simon2003}, which relies on the quantum covariance matrix. (ii) Following our approach, it seems feasible to define classical analogs of other quantum quantities, for instance, generalized purities and entropies \cite{dggv21}. (iii) It would be interesting if we could generalize these classical analogs to other kinds of states. (iv)
 In \cite{Santhanam2008} it was shown that for a classical nonlinear system the quantum entanglement of localized states is influenced by the classical bifurcations in the underlying channel periodic orbit. Also, in  \cite{Bianchi2018, Silvia2020} it was found that, under some circumstances, the linear growth of the entanglement entropy has a classical counterpart and was connected to classical and quantum quantifiers of chaos. In this sense, it would be worth to connect these results to our approach and extend our classical analogs to non-integrable systems.

\begin{acknowledgments}
Bogar D\'iaz acknowledges support from the CONEX-Plus programme funded by Universidad Carlos III de Madrid and the European Union's Horizon 2020 research and innovation program under the Marie Sklodowska-Curie Grant Agreement No. 801538. This work was partially supported by DGAPA-PAPIIT Grant No. IN105422.
\end{acknowledgments}

%
%
\appendix

\section{Proof of Eq. \eqref{rfscm}}\label{apeA}
In this appendix, we prove the relation between the Fubini-Study metric and the quantum covariance matrix, i.e., equations \eqref{rfscm}. We begin by going to the Schrödinger picture in (\ref{eq:defQGT}) to factor out the time-dependence. Analyzing the second term in (\ref{eq:defQGT}), we clearly see that $\langle\boldsymbol{\hat{\mathcal{O}}}_{A}(t_{1})\rangle_{m}=\langle m|e^{\frac{i}{\hbar}\hat{H}t_{1}}\boldsymbol{\hat{\mathcal{O}}}_{A}e^{-\frac{i}{\hbar}\hat{H}t_{1}}|m\rangle=\langle \boldsymbol{\hat{\mathcal{O}}}_{A}\rangle_{m}$, and similarly for $\langle\boldsymbol{\hat{\mathcal{O}}}_{B}(t_{2})\rangle_{m}$. Analogously, we find that the first term of (\ref{eq:defQGT}) takes the form
\begin{widetext}
\begin{equation}
\langle\boldsymbol{\hat{\mathcal{O}}}_{A}(t_{1})\boldsymbol{\hat{\mathcal{O}}}_{B}(t_{2})\rangle_{m}=e^{-\frac{i}{\hbar}E_{m}(t_{2}-t_{1})}\langle m|\boldsymbol{\hat{\mathcal{O}}}_{A}e^{-\frac{i}{\hbar}\hat{H}t_{1}}e^{\frac{i}{\hbar}\hat{H}t_{2}}\boldsymbol{\hat{\mathcal{O}}}_{B}|m\rangle\,.
\end{equation}
We introduce the identity operator $\mathbb{I}=\sum_{n}|n\rangle\langle n|$ between the two exponentials inside the bracket, which results in
\begin{equation}
\langle\boldsymbol{\hat{\mathcal{O}}}_{A}(t_{1})\boldsymbol{\hat{\mathcal{O}}}(t_{2})\rangle_{m}=\sum_{n}e^{\frac{i}{\hbar}(E_{n}-E_{m})(t_{2}-t_{1})}\langle m|\boldsymbol{\hat{\mathcal{O}}}_{A}|n\rangle\langle n|\boldsymbol{\hat{\mathcal{O}}}_{B}|m\rangle \,.
\end{equation}
Splitting the sum for $m \neq n$ and $m=n$, it is easily seen that the integrand in (\ref{eq:defQGT}) reduces to
\begin{equation}
\langle\boldsymbol{\hat{\mathcal{O}}}_{A}(t_{1})\boldsymbol{\hat{\mathcal{O}}}_{B}(t_{2})\rangle_{m}-\langle\boldsymbol{\hat{\mathcal{O}}}_{A}(t_{1})\rangle_{m}\langle\boldsymbol{\hat{\mathcal{O}}}_{B}(t_{2})\rangle_{m}=\sum_{n\neq m}e^{\frac{i}{\hbar}(E_{n}-E_{m})(t_{2}-t_{1})}\langle m|\boldsymbol{\hat{\mathcal{O}}}_{A}|n\rangle\langle n|\boldsymbol{\hat{\mathcal{O}}}_{B}|m\rangle \,,
\end{equation}
and, therefore,
\begin{equation}
G_{AB}^{(m)}=-\frac{1}{\hbar^{2}}\sum_{n\neq m}\left(\int_{-\infty}^{0} \!\!\!\mathrm{d}t_{1}\int_{0}^{\infty} \!\!\mathrm{d}t_{2}\,e^{\frac{i}{\hbar}(E_{n}-E_{m})(t_{2}-t_{1})}\right)\langle m|\boldsymbol{\hat{\mathcal{O}}}_{A}|n\rangle\langle n|\boldsymbol{\hat{\mathcal{O}}}_{B}|m\rangle\,.
\end{equation}

We thus completely isolated the time dependence and we can integrate it. Having in mind the ranges of $t_{1}$ and $t_{2}$, the computation of the integrals gives
\begin{equation}
\int_{-\infty}^{0} \!\! \mathrm{d}t_{1}\int_{0}^{\infty} \!\!\mathrm{d}t_{2}\,e^{\frac{i}{\hbar}(E_{n}-E_{m})(t_{2}-t_{1})}=\lim_{\epsilon\rightarrow0^{+}}\int_{-\infty}^{0} \!\!\mathrm{d}t_{1}\int_{0}^{\infty} \!\!\mathrm{d}t_{2}\,e^{\frac{i}{\hbar}(E_{n}-E_{m}+i\epsilon)(t_{2}-t_{1})}=-\frac{\hbar^{2}}{(E_{n}-E_{m})^{2}}\,,
\end{equation}
\end{widetext}

which yields
\begin{equation}
G_{AB}^{(m)}=\sum_{n\neq m}\frac{\langle m|\boldsymbol{\hat{\mathcal{O}}}_{A}|n\rangle\langle n|\boldsymbol{\hat{\mathcal{O}}}_{B}|m\rangle}{(E_{n}-E_{m})^{2}}\,,
\end{equation}
or specializing to phase space,
\begin{equation}
G_{\alpha \beta}^{(m)}=\sum_{n\neq m}\frac{\langle m|\boldsymbol{\hat{\mathcal{O}}}_{\alpha}|n\rangle\langle n|\boldsymbol{\hat{\mathcal{O}}}_{\beta}|m\rangle}{(E_{n}-E_{m})^{2}}\,.
\end{equation}

We first consider $\hat{\bf{q}}_a$. The Schrödinger equation reads, in position space, as
\begin{equation}
\hat{\bf{H}} \left( q,-i\hbar\frac{\partial}{\partial q} \right) \psi_{n}(q) = E_{n}\psi_{n}(q)\,,
\end{equation}
with $\psi_{n}(q)=\langle q|n \rangle $. Applying the operator $-i\hbar\frac{\partial}{\partial q_a}$, multiplying by the wave function $\psi_{m}^{*}(q)$, and integrating in $q$, we find that
\begin{align}
&-i\hbar \int \!\! \mathrm{d}^N q\, \psi_{m}^{*}(q)\frac{\partial \hat{\bf{H}}}{\partial q_a}\psi_{n}(q)  \nonumber\\
&\hspace{0.35cm} = (E_{n} - E_{m}) \int \!\! \mathrm{d}^N q\, \psi_{m}^{*}(q)\left(-i\hbar\frac{\partial \psi_{n}(q)}{\partial q_a}\right) \,,
\end{align}
or,
\begin{equation}
-i\hbar \langle m| \boldsymbol{\hat{\mathcal{O}}}_{q_a} |n \rangle = (E_{n} - E_{m}) \langle m| \hat{\bf{p}}_a |n \rangle \,. 
\end{equation}
With this result in hand, we conclude that
\begin{align}
G_{q_{a}q_{b}}^{(m)}&=\sum_{n\neq m}\frac{\langle m|\boldsymbol{\hat{\mathcal{O}}}_{q_a}|n\rangle\langle n|\boldsymbol{\hat{\mathcal{O}}}_{q_b}|m\rangle}{(E_{n}-E_{m})^{2}} \nonumber\\
&=\sum_{n\neq m}\frac{1}{\hbar^2}\langle m| \hat{\bf{p}}_a |n \rangle \langle n| \hat{\bf{p}}_b |m \rangle\,.
\end{align}
Adding and subtracting the term with $n=m$ and identifying the completeness relation, we end up with the following expression:
\begin{equation}
G_{q_{a}q_{b}}^{(m)}=\frac{1}{\hbar^2}\big{(} \langle \hat{\bf{p}}_{a} \hat{\bf{p}}_{b} \rangle_{m} - \langle \hat{\bf{p}}_{a} \rangle_{m} \langle \hat{\bf{p}}_{b} \rangle_{m} \big{)}\,. \label{gtqq}
\end{equation}
Now, we take the real part of \eqref{gtqq} to find the Fubini-Study metric for position coordinates, which turns out to be
\begin{equation}
g_{q_{a}q_{b}}^{(m)}=\frac{1}{\hbar^2}\left( \frac{1}{2}\langle \hat{\bf{p}}_{a} \hat{\bf{p}}_{b}+\hat{\bf{p}}_{b} \hat{\bf{p}}_{a} \rangle_{m} - \langle \hat{\bf{p}}_{a} \rangle_{m} \langle \hat{\bf{p}}_{b} \rangle_{m} \right)\,.
\end{equation}

On the other hand, writing the Schrödinger equation in momentum space, we see that the following relation holds for $p_a$:
\begin{equation}
i\hbar \langle m| \boldsymbol{\hat{\mathcal{O}}}_{p_a} |n \rangle = (E_{n} - E_{m}) \langle m| \hat{\bf{q}}_a |n \rangle \,. 
\end{equation}
This allows us to proceed similarly with $g_{q_{a}p_{b}}^{(m)}$ and $g_{p_{a}p_{b}}^{(m)}$ to find that
\begin{equation}
g_{q_{a}p_{b}}^{(m)}=-\frac{1}{\hbar^2}\left( \frac{1}{2}\langle \hat{\bf{p}}_{a} \hat{\bf{q}}_{b}+\hat{\bf{q}}_{b} \hat{\bf{p}}_{a} \rangle_{m} - \langle \hat{\bf{p}}_{a} \rangle_{m} \langle \hat{\bf{q}}_{b} \rangle_{m} \right)\,,
\end{equation}
and
\begin{equation}
g_{p_{a}p_{b}}^{(m)}=\frac{1}{\hbar^2}\left( \frac{1}{2}\langle \hat{\bf{q}}_{a} \hat{\bf{q}}_{b}+\hat{\bf{q}}_{b} \hat{\bf{q}}_{a} \rangle_{m} - \langle \hat{\bf{q}}_{a} \rangle_{m} \langle \hat{\bf{q}}_{b} \rangle_{m} \right)\,.
\end{equation}
This completes the proof.
\vspace{0.5cm}

\section{On the classical average}\label{apenB}

In this appendix, we provide another perspective on the classical average, we introduce an angular probability distribution $P_{\mathrm{cl}}(\varphi)$ (see \cite{Robinett,Gattus} for the usual position-space treatment). Without loss of generality, we consider a classical system with one degree of freedom and the corresponding orbit in phase space for a fixed energy. The probability of finding the particle in the orbit's region spanned by $\mathrm{d} \varphi$ is given by the ratio of time $\mathrm{d} t$ that it spends there, and the period $T=2\pi/\omega$ that it takes to traverse the entire orbit. More precisely, this probability is given by 
\begin{equation}
P_{\mathrm{cl}}(\varphi) \mathrm{d}\varphi := \frac{\mathrm{d} t}{T} = \frac{\mathrm{d} \varphi}{\omega T} = \frac{\mathrm{d} \varphi}{2\pi}\,.
\end{equation}
From this expression, we read off the  probability distribution $P_{\mathrm{cl}}(\varphi)=1/(2\pi)$, which is uniform and allows us to compute the average $\bar{f}(I)$ of a function $f(\varphi,I)$ as 
\begin{equation}
\bar{f} := \int_{0}^{2\pi} \!\! \mathrm{d} \varphi \, P_{\mathrm{cl}}(\varphi) f(\varphi) = \langle f \rangle_{\mathrm{cl}}\,,
\end{equation}
matching the  classical average (\ref{classAvg}). An analogous result follows for integrable systems with $N$ degrees of freedom. If the system is separable and the analysis is made in each degree of freedom's phase space $(q_{a},p_{a})$, the total probability distribution turns out to be
\begin{equation}
P_{\mathrm{cl}}(\varphi^{1},...,\varphi^{N}) = P_{\mathrm{cl}}^{(1)}(\varphi^{1}) \cdots P_{\mathrm{cl}}^{(N)}(\varphi^{N})\,.
\end{equation}
The same result follows for integrable but not separable systems (in the original coordinates $(q,p)$), once transformed into normal coordinates.

\bibliography{journalpra}

\begin{thebibliography}{56}%
\makeatletter
\providecommand \@ifxundefined [1]{%
 \@ifx{#1\undefined}
}%
\providecommand \@ifnum [1]{%
 \ifnum #1\expandafter \@firstoftwo
 \else \expandafter \@secondoftwo
 \fi
}%
\providecommand \@ifx [1]{%
 \ifx #1\expandafter \@firstoftwo
 \else \expandafter \@secondoftwo
 \fi
}%
\providecommand \natexlab [1]{#1}%
\providecommand \enquote  [1]{``#1''}%
\providecommand \bibnamefont  [1]{#1}%
\providecommand \bibfnamefont [1]{#1}%
\providecommand \citenamefont [1]{#1}%
\providecommand \href@noop [0]{\@secondoftwo}%
\providecommand \href [0]{\begingroup \@sanitize@url \@href}%
\providecommand \@href[1]{\@@startlink{#1}\@@href}%
\providecommand \@@href[1]{\endgroup#1\@@endlink}%
\providecommand \@sanitize@url [0]{\catcode `\\12\catcode `\$12\catcode
  `\&12\catcode `\#12\catcode `\^12\catcode `\_12\catcode `\%12\relax}%
\providecommand \@@startlink[1]{}%
\providecommand \@@endlink[0]{}%
\providecommand \url  [0]{\begingroup\@sanitize@url \@url }%
\providecommand \@url [1]{\endgroup\@href {#1}{\urlprefix }}%
\providecommand \urlprefix  [0]{URL }%
\providecommand \Eprint [0]{\href }%
\providecommand \doibase [0]{https://doi.org/}%
\providecommand \selectlanguage [0]{\@gobble}%
\providecommand \bibinfo  [0]{\@secondoftwo}%
\providecommand \bibfield  [0]{\@secondoftwo}%
\providecommand \translation [1]{[#1]}%
\providecommand \BibitemOpen [0]{}%
\providecommand \bibitemStop [0]{}%
\providecommand \bibitemNoStop [0]{.\EOS\space}%
\providecommand \EOS [0]{\spacefactor3000\relax}%
\providecommand \BibitemShut  [1]{\csname bibitem#1\endcsname}%
\let\auto@bib@innerbib\@empty
\bibitem [{\citenamefont {Rokhlin}(1967)}]{Rokhlin_1967}%
  \BibitemOpen
  \bibfield  {author} {\bibinfo {author} {\bibfnamefont {V.~A.}\ \bibnamefont
  {Rokhlin}},\ }\bibfield  {title} {\bibinfo {title} {Lectures on the entropy
  theory of measure-preserving transformations},\ }\href
  {https://doi.org/10.1070/rm1967v022n05abeh001224} {\bibfield  {journal}
  {\bibinfo  {journal} {Russian Mathematical Surveys}\ }\textbf {\bibinfo
  {volume} {22}},\ \bibinfo {pages} {1} (\bibinfo {year} {1967})}\BibitemShut
  {NoStop}%
\bibitem [{\citenamefont {Rajski}(1961)}]{Rajski1961}%
  \BibitemOpen
  \bibfield  {author} {\bibinfo {author} {\bibfnamefont {C.}~\bibnamefont
  {Rajski}},\ }\bibfield  {title} {\bibinfo {title} {A metric space of discrete
  probability distributions},\ }\href
  {https://doi.org/10.1016/S0019-9958(61)80055-7} {\bibfield  {journal}
  {\bibinfo  {journal} {Information and Control}\ }\textbf {\bibinfo {volume}
  {4}},\ \bibinfo {pages} {371} (\bibinfo {year} {1961})}\BibitemShut {NoStop}%
\bibitem [{\citenamefont {Schumacher}(1991)}]{Schumacher1991}%
  \BibitemOpen
  \bibfield  {author} {\bibinfo {author} {\bibfnamefont {B.~W.}\ \bibnamefont
  {Schumacher}},\ }\bibfield  {title} {\bibinfo {title} {Information and
  quantum nonseparability},\ }\href {https://doi.org/10.1103/PhysRevA.44.7047}
  {\bibfield  {journal} {\bibinfo  {journal} {Phys. Rev. A}\ }\textbf {\bibinfo
  {volume} {44}},\ \bibinfo {pages} {7047} (\bibinfo {year}
  {1991})}\BibitemShut {NoStop}%
\bibitem [{\citenamefont {Uhlmann}(1976)}]{Uhlmann1976}%
  \BibitemOpen
  \bibfield  {author} {\bibinfo {author} {\bibfnamefont {A.}~\bibnamefont
  {Uhlmann}},\ }\bibfield  {title} {\bibinfo {title} {The “transition
  probability” in the state space of a $*$-algebra},\ }\href
  {https://doi.org/10.1016/0034-4877(76)90060-4} {\bibfield  {journal}
  {\bibinfo  {journal} {Reports on Mathematical Physics}\ }\textbf {\bibinfo
  {volume} {9}},\ \bibinfo {pages} {273} (\bibinfo {year} {1976})}\BibitemShut
  {NoStop}%
\bibitem [{\citenamefont {Provost}\ and\ \citenamefont
  {Vallee}(1980)}]{Provost1980}%
  \BibitemOpen
  \bibfield  {author} {\bibinfo {author} {\bibfnamefont {J.~P.}\ \bibnamefont
  {Provost}}\ and\ \bibinfo {author} {\bibfnamefont {G.}~\bibnamefont
  {Vallee}},\ }\bibfield  {title} {\bibinfo {title} {Riemannian structure on
  manifolds of quantum states},\ }\href {https://doi.org/10.1007/BF02193559}
  {\bibfield  {journal} {\bibinfo  {journal} {Commun. Math. Phys.}\ }\textbf
  {\bibinfo {volume} {76}},\ \bibinfo {pages} {289} (\bibinfo {year}
  {1980})}\BibitemShut {NoStop}%
\bibitem [{\citenamefont {Berry}(1984)}]{Berry1984}%
  \BibitemOpen
  \bibfield  {author} {\bibinfo {author} {\bibfnamefont {M.~V.}\ \bibnamefont
  {Berry}},\ }\bibfield  {title} {\bibinfo {title} {Quantal phase factors
  accompanying adiabatic changes},\ }\href
  {https://doi.org/10.1098/rspa.1984.0023} {\bibfield  {journal} {\bibinfo
  {journal} {Proce Royal Soc. A: Math. Phys. Sci.}\ }\textbf {\bibinfo {volume}
  {392}},\ \bibinfo {pages} {45} (\bibinfo {year} {1984})}\BibitemShut
  {NoStop}%
\bibitem [{\citenamefont {Gorin}\ \emph {et~al.}(2006)\citenamefont {Gorin},
  \citenamefont {Prosen}, \citenamefont {Seligman},\ and\ \citenamefont
  {\v{Z}nidari\v{c}}}]{Gorin}%
  \BibitemOpen
  \bibfield  {author} {\bibinfo {author} {\bibfnamefont {T.}~\bibnamefont
  {Gorin}}, \bibinfo {author} {\bibfnamefont {T.}~\bibnamefont {Prosen}},
  \bibinfo {author} {\bibfnamefont {T.~H.}\ \bibnamefont {Seligman}},\ and\
  \bibinfo {author} {\bibfnamefont {M.}~\bibnamefont {\v{Z}nidari\v{c}}},\
  }\bibfield  {title} {\bibinfo {title} {Dynamics of loschmidt echoes and
  fidelity decay},\ }\href {https://doi.org/10.1016/j.physrep.2006.09.003}
  {\bibfield  {journal} {\bibinfo  {journal} {Physics Reports}\ }\textbf
  {\bibinfo {volume} {435}},\ \bibinfo {pages} {33} (\bibinfo {year}
  {2006})}\BibitemShut {NoStop}%
\bibitem [{\citenamefont {Alvarez-Jimenez}\ \emph {et~al.}(2017)\citenamefont
  {Alvarez-Jimenez}, \citenamefont {Dector},\ and\ \citenamefont
  {Vergara}}]{ADV2017}%
  \BibitemOpen
  \bibfield  {author} {\bibinfo {author} {\bibfnamefont {J.}~\bibnamefont
  {Alvarez-Jimenez}}, \bibinfo {author} {\bibfnamefont {A.}~\bibnamefont
  {Dector}},\ and\ \bibinfo {author} {\bibfnamefont {J.~D.}\ \bibnamefont
  {Vergara}},\ }\bibfield  {title} {\bibinfo {title} {Quantum information
  metric and {B}erry curvature from a {L}agrangian approach},\ }\href
  {https://doi.org/10.1007/JHEP03(2017)044} {\bibfield  {journal} {\bibinfo
  {journal} {Journal of High Energy Physics}\ }\textbf {\bibinfo {volume}
  {2017}},\ \bibinfo {pages} {44} (\bibinfo {year} {2017})}\BibitemShut
  {NoStop}%
\bibitem [{\citenamefont {Hannay}(1985)}]{Hannay_1985}%
  \BibitemOpen
  \bibfield  {author} {\bibinfo {author} {\bibfnamefont {J.~H.}\ \bibnamefont
  {Hannay}},\ }\bibfield  {title} {\bibinfo {title} {Angle variable holonomy in
  adiabatic excursion of an integrable {H}amiltonian},\ }\href
  {https://doi.org/10.1088/0305-4470/18/2/011} {\bibfield  {journal} {\bibinfo
  {journal} {Journal of Physics A: Mathematical and General}\ }\textbf
  {\bibinfo {volume} {18}},\ \bibinfo {pages} {221} (\bibinfo {year}
  {1985})}\BibitemShut {NoStop}%
\bibitem [{\citenamefont {Chaturvedi}\ \emph {et~al.}(1987)\citenamefont
  {Chaturvedi}, \citenamefont {Sriram},\ and\ \citenamefont
  {Srinivasan}}]{Chaturvedi1987}%
  \BibitemOpen
  \bibfield  {author} {\bibinfo {author} {\bibfnamefont {S.}~\bibnamefont
  {Chaturvedi}}, \bibinfo {author} {\bibfnamefont {M.~S.}\ \bibnamefont
  {Sriram}},\ and\ \bibinfo {author} {\bibfnamefont {V.}~\bibnamefont
  {Srinivasan}},\ }\bibfield  {title} {\bibinfo {title}
  {Berry{\textquotesingle}s phase for coherent states},\ }\href
  {https://doi.org/10.1088/0305-4470/20/16/007} {\bibfield  {journal} {\bibinfo
   {journal} {Journal of Physics A: Mathematical and General}\ }\textbf
  {\bibinfo {volume} {20}},\ \bibinfo {pages} {L1071} (\bibinfo {year}
  {1987})}\BibitemShut {NoStop}%
\bibitem [{\citenamefont {Yong-de}\ and\ \citenamefont {Lei}(1990)}]{Zhang}%
  \BibitemOpen
  \bibfield  {author} {\bibinfo {author} {\bibfnamefont {Z.}~\bibnamefont
  {Yong-de}}\ and\ \bibinfo {author} {\bibfnamefont {M.}~\bibnamefont {Lei}},\
  }\bibfield  {title} {\bibinfo {title} {Berry's phase for coherent states},\
  }\href {https://doi.org/10.1007/BF02742688} {\bibfield  {journal} {\bibinfo
  {journal} {Nuov Cim B}\ }\textbf {\bibinfo {volume} {105}},\ \bibinfo {pages}
  {1343} (\bibinfo {year} {1990})}\BibitemShut {NoStop}%
\bibitem [{\citenamefont {Gonzalez}\ \emph {et~al.}(2019)\citenamefont
  {Gonzalez}, \citenamefont {Guti\'errez-Ruiz},\ and\ \citenamefont
  {Vergara}}]{Gonzales2019}%
  \BibitemOpen
  \bibfield  {author} {\bibinfo {author} {\bibfnamefont {D.}~\bibnamefont
  {Gonzalez}}, \bibinfo {author} {\bibfnamefont {D.}~\bibnamefont
  {Guti\'errez-Ruiz}},\ and\ \bibinfo {author} {\bibfnamefont {J.~D.}\
  \bibnamefont {Vergara}},\ }\bibfield  {title} {\bibinfo {title} {Classical
  analog of the quantum metric tensor},\ }\href
  {https://doi.org/10.1103/PhysRevE.99.032144} {\bibfield  {journal} {\bibinfo
  {journal} {Phys. Rev. E}\ }\textbf {\bibinfo {volume} {99}},\ \bibinfo
  {pages} {032144} (\bibinfo {year} {2019})}\BibitemShut {NoStop}%
\bibitem [{\citenamefont {Alvarez-Jimenez}\ \emph {et~al.}(2020)\citenamefont
  {Alvarez-Jimenez}, \citenamefont {Gonzalez}, \citenamefont
  {Gutiérrez-Ruiz},\ and\ \citenamefont {Vergara}}]{Alvarez2019}%
  \BibitemOpen
  \bibfield  {author} {\bibinfo {author} {\bibfnamefont {J.}~\bibnamefont
  {Alvarez-Jimenez}}, \bibinfo {author} {\bibfnamefont {D.}~\bibnamefont
  {Gonzalez}}, \bibinfo {author} {\bibfnamefont {D.}~\bibnamefont
  {Gutiérrez-Ruiz}},\ and\ \bibinfo {author} {\bibfnamefont {J.~D.}\
  \bibnamefont {Vergara}},\ }\bibfield  {title} {\bibinfo {title} {Geometry of
  the parameter space of a quantum system: Classical point of view},\ }\href
  {https://doi.org/10.1002/andp.201900215} {\bibfield  {journal} {\bibinfo
  {journal} {Ann. Phys. (Leipzig)}\ }\textbf {\bibinfo {volume} {532}},\
  \bibinfo {pages} {1900215} (\bibinfo {year} {2020})}\BibitemShut {NoStop}%
\bibitem [{\citenamefont {Gonzalez}\ \emph {et~al.}(2020)\citenamefont
  {Gonzalez}, \citenamefont {Guti{\'{e}}rrez-Ruiz},\ and\ \citenamefont
  {Vergara}}]{Gonzalez2020}%
  \BibitemOpen
  \bibfield  {author} {\bibinfo {author} {\bibfnamefont {D.}~\bibnamefont
  {Gonzalez}}, \bibinfo {author} {\bibfnamefont {D.}~\bibnamefont
  {Guti{\'{e}}rrez-Ruiz}},\ and\ \bibinfo {author} {\bibfnamefont {J.~D.}\
  \bibnamefont {Vergara}},\ }\bibfield  {title} {\bibinfo {title} {Phase space
  formulation of the {A}belian and non-{A}belian quantum geometric tensor},\
  }\href {https://doi.org/10.1088/1751-8121/abc6c2} {\bibfield  {journal}
  {\bibinfo  {journal} {Journal of Physics A: Mathematical and Theoretical}\
  }\textbf {\bibinfo {volume} {53}},\ \bibinfo {pages} {505305} (\bibinfo
  {year} {2020})}\BibitemShut {NoStop}%
\bibitem [{\citenamefont {Bell}\ and\ \citenamefont {Aspect}(2004)}]{Bell}%
  \BibitemOpen
  \bibfield  {author} {\bibinfo {author} {\bibfnamefont {J.}~\bibnamefont
  {Bell}}\ and\ \bibinfo {author} {\bibfnamefont {A.}~\bibnamefont {Aspect}},\
  }\href@noop {} {\emph {\bibinfo {title} {Speakable and Unspeakable in Quantum
  Mechanics: Collected Papers on Quantum Philosophy}}},\ Collected papers on
  quantum philosophy,\ (\bibinfo  {publisher} {Cambridge University Press},\
  \bibinfo {address} {Cambridge, England},\ \bibinfo {year} {2004})\BibitemShut
  {NoStop}%
\bibitem [{\citenamefont {Paneru}\ \emph {et~al.}(2020)\citenamefont {Paneru},
  \citenamefont {Cohen}, \citenamefont {Fickler}, \citenamefont {Boyd},\ and\
  \citenamefont {Karimi}}]{Paneru_2020}%
  \BibitemOpen
  \bibfield  {author} {\bibinfo {author} {\bibfnamefont {D.}~\bibnamefont
  {Paneru}}, \bibinfo {author} {\bibfnamefont {E.}~\bibnamefont {Cohen}},
  \bibinfo {author} {\bibfnamefont {R.}~\bibnamefont {Fickler}}, \bibinfo
  {author} {\bibfnamefont {R.~W.}\ \bibnamefont {Boyd}},\ and\ \bibinfo
  {author} {\bibfnamefont {E.}~\bibnamefont {Karimi}},\ }\bibfield  {title}
  {\bibinfo {title} {Entanglement: quantum or classical?},\ }\href
  {https://doi.org/10.1088/1361-6633/ab85b9} {\bibfield  {journal} {\bibinfo
  {journal} {Reports on Progress in Physics}\ }\textbf {\bibinfo {volume}
  {83}},\ \bibinfo {pages} {064001} (\bibinfo {year} {2020})}\BibitemShut
  {NoStop}%
\bibitem [{\citenamefont {Gong}\ and\ \citenamefont {Brumer}(2003)}]{Gong2003}%
  \BibitemOpen
  \bibfield  {author} {\bibinfo {author} {\bibfnamefont {J.}~\bibnamefont
  {Gong}}\ and\ \bibinfo {author} {\bibfnamefont {P.}~\bibnamefont {Brumer}},\
  }\bibfield  {title} {\bibinfo {title} {Intrinsic decoherence dynamics in
  smooth hamiltonian systems: Quantum-classical correspondence},\ }\href
  {https://doi.org/10.1103/PhysRevA.68.022101} {\bibfield  {journal} {\bibinfo
  {journal} {Phys. Rev. A}\ }\textbf {\bibinfo {volume} {68}},\ \bibinfo
  {pages} {022101} (\bibinfo {year} {2003})}\BibitemShut {NoStop}%
\bibitem [{\citenamefont {Gittsovich}\ and\ \citenamefont
  {G\"uhne}(2010)}]{Gittsovich2010}%
  \BibitemOpen
  \bibfield  {author} {\bibinfo {author} {\bibfnamefont {O.}~\bibnamefont
  {Gittsovich}}\ and\ \bibinfo {author} {\bibfnamefont {O.}~\bibnamefont
  {G\"uhne}},\ }\bibfield  {title} {\bibinfo {title} {Quantifying entanglement
  with covariance matrices},\ }\href
  {https://doi.org/10.1103/PhysRevA.81.032333} {\bibfield  {journal} {\bibinfo
  {journal} {Phys. Rev. A}\ }\textbf {\bibinfo {volume} {81}},\ \bibinfo
  {pages} {032333} (\bibinfo {year} {2010})}\BibitemShut {NoStop}%
\bibitem [{\citenamefont {Furusawa}\ \emph {et~al.}(1998)\citenamefont
  {Furusawa}, \citenamefont {S{\o}rensen}, \citenamefont {Braunstein},
  \citenamefont {Fuchs}, \citenamefont {Kimble},\ and\ \citenamefont
  {Polzik}}]{Furusawa}%
  \BibitemOpen
  \bibfield  {author} {\bibinfo {author} {\bibfnamefont {A.}~\bibnamefont
  {Furusawa}}, \bibinfo {author} {\bibfnamefont {J.~L.}\ \bibnamefont
  {S{\o}rensen}}, \bibinfo {author} {\bibfnamefont {S.~L.}\ \bibnamefont
  {Braunstein}}, \bibinfo {author} {\bibfnamefont {C.~A.}\ \bibnamefont
  {Fuchs}}, \bibinfo {author} {\bibfnamefont {H.~J.}\ \bibnamefont {Kimble}},\
  and\ \bibinfo {author} {\bibfnamefont {E.~S.}\ \bibnamefont {Polzik}},\
  }\bibfield  {title} {\bibinfo {title} {Unconditional quantum teleportation},\
  }\href {https://doi.org/10.1126/science.282.5389.706} {\bibfield  {journal}
  {\bibinfo  {journal} {Science}\ }\textbf {\bibinfo {volume} {282}},\ \bibinfo
  {pages} {706} (\bibinfo {year} {1998})}\BibitemShut {NoStop}%
\bibitem [{\citenamefont {Jeong}\ \emph {et~al.}(2007)\citenamefont {Jeong},
  \citenamefont {Ralph},\ and\ \citenamefont {Bowen}}]{Jeong}%
  \BibitemOpen
  \bibfield  {author} {\bibinfo {author} {\bibfnamefont {H.}~\bibnamefont
  {Jeong}}, \bibinfo {author} {\bibfnamefont {T.~C.}\ \bibnamefont {Ralph}},\
  and\ \bibinfo {author} {\bibfnamefont {W.~P.}\ \bibnamefont {Bowen}},\
  }\bibfield  {title} {\bibinfo {title} {Quantum and classical fidelities for
  {G}aussian states},\ }\href {https://doi.org/10.1364/JOSAB.24.000355}
  {\bibfield  {journal} {\bibinfo  {journal} {J. Opt. Soc. Am. B}\ }\textbf
  {\bibinfo {volume} {24}},\ \bibinfo {pages} {355} (\bibinfo {year}
  {2007})}\BibitemShut {NoStop}%
\bibitem [{\citenamefont {Collins}\ and\ \citenamefont
  {Popescu}(2002)}]{Collins-Popescu}%
  \BibitemOpen
  \bibfield  {author} {\bibinfo {author} {\bibfnamefont {D.}~\bibnamefont
  {Collins}}\ and\ \bibinfo {author} {\bibfnamefont {S.}~\bibnamefont
  {Popescu}},\ }\bibfield  {title} {\bibinfo {title} {Classical analog of
  entanglement},\ }\href {https://doi.org/10.1103/PhysRevA.65.032321}
  {\bibfield  {journal} {\bibinfo  {journal} {Phys. Rev. A}\ }\textbf {\bibinfo
  {volume} {65}},\ \bibinfo {pages} {032321} (\bibinfo {year}
  {2002})}\BibitemShut {NoStop}%
\bibitem [{\citenamefont {Abe}(1993)}]{Abe1993}%
  \BibitemOpen
  \bibfield  {author} {\bibinfo {author} {\bibfnamefont {S.}~\bibnamefont
  {Abe}},\ }\bibfield  {title} {\bibinfo {title} {Quantized geometry associated
  with uncertainty and correlation},\ }\href
  {https://doi.org/10.1103/PhysRevA.48.4102} {\bibfield  {journal} {\bibinfo
  {journal} {Phys. Rev. A}\ }\textbf {\bibinfo {volume} {48}},\ \bibinfo
  {pages} {4102} (\bibinfo {year} {1993})}\BibitemShut {NoStop}%
\bibitem [{\citenamefont {Wigner}(1932)}]{Wigner1932}%
  \BibitemOpen
  \bibfield  {author} {\bibinfo {author} {\bibfnamefont {E.}~\bibnamefont
  {Wigner}},\ }\bibfield  {title} {\bibinfo {title} {On the quantum correction
  for thermodynamic equilibrium},\ }\href
  {https://doi.org/10.1103/PhysRev.40.749} {\bibfield  {journal} {\bibinfo
  {journal} {Phys. Rev.}\ }\textbf {\bibinfo {volume} {40}},\ \bibinfo {pages}
  {749} (\bibinfo {year} {1932})}\BibitemShut {NoStop}%
\bibitem [{\citenamefont {Hillery}\ \emph {et~al.}(1984)\citenamefont
  {Hillery}, \citenamefont {O'Connell}, \citenamefont {Scully},\ and\
  \citenamefont {Wigner}}]{Hillery1984}%
  \BibitemOpen
  \bibfield  {author} {\bibinfo {author} {\bibfnamefont {M.}~\bibnamefont
  {Hillery}}, \bibinfo {author} {\bibfnamefont {R.}~\bibnamefont {O'Connell}},
  \bibinfo {author} {\bibfnamefont {M.}~\bibnamefont {Scully}},\ and\ \bibinfo
  {author} {\bibfnamefont {E.}~\bibnamefont {Wigner}},\ }\bibfield  {title}
  {\bibinfo {title} {Distribution functions in physics: Fundamentals},\ }\href
  {https://doi.org/10.1016/0370-1573(84)90160-1} {\bibfield  {journal}
  {\bibinfo  {journal} {Physics Reports}\ }\textbf {\bibinfo {volume} {106}},\
  \bibinfo {pages} {121} (\bibinfo {year} {1984})}\BibitemShut {NoStop}%
\bibitem [{\citenamefont {Case}(2008)}]{Case2008}%
  \BibitemOpen
  \bibfield  {author} {\bibinfo {author} {\bibfnamefont {W.~B.}\ \bibnamefont
  {Case}},\ }\bibfield  {title} {\bibinfo {title} {Wigner functions and {W}eyl
  transforms for pedestrians},\ }\href {https://doi.org/10.1119/1.2957889}
  {\bibfield  {journal} {\bibinfo  {journal} {American Journal of Physics}\
  }\textbf {\bibinfo {volume} {76}},\ \bibinfo {pages} {937} (\bibinfo {year}
  {2008})}\BibitemShut {NoStop}%
\bibitem [{\citenamefont {Berry}(1977)}]{Berry1977}%
  \BibitemOpen
  \bibfield  {author} {\bibinfo {author} {\bibfnamefont {M.~V.}\ \bibnamefont
  {Berry}},\ }\bibfield  {title} {\bibinfo {title} {Semi-classical mechanics in
  phase space: A study of {W}igner's function},\ }\href
  {https://doi.org/10.1098/rsta.1977.0145} {\bibfield  {journal} {\bibinfo
  {journal} {Philosophical Transactions of the Royal Society of London. Series
  A, Mathematical and Physical Sciences}\ }\textbf {\bibinfo {volume} {287}},\
  \bibinfo {pages} {237} (\bibinfo {year} {1977})}\BibitemShut {NoStop}%
\bibitem [{\citenamefont {Zachos}\ \emph {et~al.}(2005)\citenamefont {Zachos},
  \citenamefont {Fairlie},\ and\ \citenamefont {Curtright}}]{Zachos}%
  \BibitemOpen
  \bibfield  {author} {\bibinfo {author} {\bibfnamefont {C.~K.}\ \bibnamefont
  {Zachos}}, \bibinfo {author} {\bibfnamefont {D.~B.}\ \bibnamefont
  {Fairlie}},\ and\ \bibinfo {author} {\bibfnamefont {T.~L.}\ \bibnamefont
  {Curtright}},\ }\href {https://doi.org/10.1142/5287} {\emph {\bibinfo {title}
  {Quantum Mechanics in Phase Space}}}\ (\bibinfo  {publisher} {World
  Scientific},\ \bibinfo {address} {Singapure},\ \bibinfo {year} {2005})\
  \Eprint
  {https://arxiv.org/abs/https://www.worldscientific.com/doi/pdf/10.1142/5287}
  {https://www.worldscientific.com/doi/pdf/10.1142/5287} \BibitemShut {NoStop}%
\bibitem [{\citenamefont {Huang}\ and\ \citenamefont
  {Huang}(2011)}]{Huang2011}%
  \BibitemOpen
  \bibfield  {author} {\bibinfo {author} {\bibfnamefont {C.}~\bibnamefont
  {Huang}}\ and\ \bibinfo {author} {\bibfnamefont {Y.-C.}\ \bibnamefont
  {Huang}},\ }\bibfield  {title} {\bibinfo {title} {Unification theory of
  classical statistical uncertainty relation and quantum uncertainty relation
  and its applications},\ }\href
  {https://doi.org/https://doi.org/10.1016/j.physleta.2010.11.005} {\bibfield
  {journal} {\bibinfo  {journal} {Physics Letters A}\ }\textbf {\bibinfo
  {volume} {375}},\ \bibinfo {pages} {271} (\bibinfo {year}
  {2011})}\BibitemShut {NoStop}%
\bibitem [{\citenamefont {Dittrich}\ and\ \citenamefont
  {Reuter}(2001)}]{Dittrich}%
  \BibitemOpen
  \bibfield  {author} {\bibinfo {author} {\bibfnamefont {W.}~\bibnamefont
  {Dittrich}}\ and\ \bibinfo {author} {\bibfnamefont {M.}~\bibnamefont
  {Reuter}},\ }\href@noop {} {\emph {\bibinfo {title} {Classical and Quantum
  Dynamics: From Classical Paths to Path Integrals}}},\ \bibinfo {edition}
  {3rd}\ ed.\ (\bibinfo  {publisher} {Springer, New York},\ \bibinfo {year}
  {2001})\BibitemShut {NoStop}%
\bibitem [{\citenamefont {Turbiner}(1984)}]{Turbiner1984}%
  \BibitemOpen
  \bibfield  {author} {\bibinfo {author} {\bibfnamefont {A.~V.}\ \bibnamefont
  {Turbiner}},\ }\bibfield  {title} {\bibinfo {title} {The eigenvalue spectrum
  in quantum mechanics and the nonlinearization procedure},\ }\href
  {https://doi.org/10.1070%2Fpu1984v027n09abeh004155} {\bibfield  {journal}
  {\bibinfo  {journal} {Soviet Physics Uspekhi}\ }\textbf {\bibinfo {volume}
  {27}},\ \bibinfo {pages} {668} (\bibinfo {year} {1984})}\BibitemShut
  {NoStop}%
\bibitem [{\citenamefont {Fl{\"{u}}gge}(1999)}]{Flugge}%
  \BibitemOpen
  \bibfield  {author} {\bibinfo {author} {\bibfnamefont {S.}~\bibnamefont
  {Fl{\"{u}}gge}},\ }\href@noop {} {\emph {\bibinfo {title} {Practical Quantum
  Mechanics}}},\ Classics in Mathematics,\ (\bibinfo  {publisher} {Springer,
  New York},\ \bibinfo {year} {1999})\BibitemShut {NoStop}%
\bibitem [{\citenamefont {Guti{\'{e}}rrez-Ruiz}(2021)}]{Thesis2021}%
  \BibitemOpen
  \bibfield  {author} {\bibinfo {author} {\bibfnamefont {D.}~\bibnamefont
  {Guti{\'{e}}rrez-Ruiz}},\ }\emph {\bibinfo {title} {Classical and quantum
  descriptions of the parameter space geometry}},\ \href@noop {} {Ph.D.
  thesis},\ \bibinfo  {school} {Universidad Nacional Aut{\'{o}}noma de
  M{\'{e}}xico}, \bibinfo {address} {Mexico City} (\bibinfo {year}
  {2021})\BibitemShut {NoStop}%
\bibitem [{\citenamefont {de~Gosson}(2006)}]{deGosson2006}%
  \BibitemOpen
  \bibfield  {author} {\bibinfo {author} {\bibfnamefont {M.}~\bibnamefont
  {de~Gosson}},\ }\href {https://doi.org/10.1007/3-7643-7575-2} {\emph
  {\bibinfo {title} {Symplectic Geometry and Quantum Mechanics}}},\ Operator
  Theory: Advances and Applications\ (\bibinfo  {publisher} {Birkh{\"a}user,
  Basel, Switzerland},\ \bibinfo {year} {2006})\BibitemShut {NoStop}%
\bibitem [{\citenamefont {Diosi}(2011)}]{diosi2011}%
  \BibitemOpen
  \bibfield  {author} {\bibinfo {author} {\bibfnamefont {L.}~\bibnamefont
  {Diosi}},\ }\href {https://books.google.es/books?id=w13vCAAAQBAJ} {\emph
  {\bibinfo {title} {A Short Course in Quantum Information Theory: An Approach
  From Theoretical Physics}}},\ Lecture Notes in Physics\ (\bibinfo
  {publisher} {Springer Berlin Heidelberg},\ \bibinfo {year}
  {2011})\BibitemShut {NoStop}%
\bibitem [{\citenamefont {Dodonov}(2002)}]{Dodonov_2002}%
  \BibitemOpen
  \bibfield  {author} {\bibinfo {author} {\bibfnamefont {V.~V.}\ \bibnamefont
  {Dodonov}},\ }\bibfield  {title} {\bibinfo {title} {Purity- and
  entropy-bounded uncertainty relations for mixed quantum states},\ }\href
  {https://doi.org/10.1088/1464-4266/4/3/362} {\bibfield  {journal} {\bibinfo
  {journal} {Journal of Optics B: Quantum and Semiclassical Optics}\ }\textbf
  {\bibinfo {volume} {4}},\ \bibinfo {pages} {S98} (\bibinfo {year}
  {2002})}\BibitemShut {NoStop}%
\bibitem [{\citenamefont {Paris}\ \emph {et~al.}(2003)\citenamefont {Paris},
  \citenamefont {Illuminati}, \citenamefont {Serafini},\ and\ \citenamefont
  {De~Siena}}]{Paris2003}%
  \BibitemOpen
  \bibfield  {author} {\bibinfo {author} {\bibfnamefont {M.~G.~A.}\
  \bibnamefont {Paris}}, \bibinfo {author} {\bibfnamefont {F.}~\bibnamefont
  {Illuminati}}, \bibinfo {author} {\bibfnamefont {A.}~\bibnamefont
  {Serafini}},\ and\ \bibinfo {author} {\bibfnamefont {S.}~\bibnamefont
  {De~Siena}},\ }\bibfield  {title} {\bibinfo {title} {Purity of {G}aussian
  states: Measurement schemes and time evolution in noisy channels},\ }\href
  {https://doi.org/10.1103/PhysRevA.68.012314} {\bibfield  {journal} {\bibinfo
  {journal} {Phys. Rev. A}\ }\textbf {\bibinfo {volume} {68}},\ \bibinfo
  {pages} {012314} (\bibinfo {year} {2003})}\BibitemShut {NoStop}%
\bibitem [{\citenamefont {Golubeva}\ and\ \citenamefont
  {Golubev}(2014)}]{Golubeva2014}%
  \BibitemOpen
  \bibfield  {author} {\bibinfo {author} {\bibfnamefont {T.}~\bibnamefont
  {Golubeva}}\ and\ \bibinfo {author} {\bibfnamefont {Y.}~\bibnamefont
  {Golubev}},\ }\bibfield  {title} {\bibinfo {title} {Purity and covariance
  matrix},\ }\href {https://doi.org/10.1007/s10946-014-9399-2} {\bibfield
  {journal} {\bibinfo  {journal} {Journal of Russian Laser Research}\ }\textbf
  {\bibinfo {volume} {35}},\ \bibinfo {pages} {47} (\bibinfo {year}
  {2014})}\BibitemShut {NoStop}%
\bibitem [{\citenamefont {Serafini}(2017)}]{serafini2017}%
  \BibitemOpen
  \bibfield  {author} {\bibinfo {author} {\bibfnamefont {A.}~\bibnamefont
  {Serafini}},\ }\href {https://books.google.es/books?id=zHtgvgAACAAJ} {\emph
  {\bibinfo {title} {Quantum Continuous Variables: A Primer of Theoretical
  Methods}}}\ (\bibinfo  {publisher} {CRC Press, Boca Raton, FL},\ \bibinfo
  {year} {2017})\BibitemShut {NoStop}%
\bibitem [{\citenamefont {de~Gosson}(2019)}]{deGosson2019}%
  \BibitemOpen
  \bibfield  {author} {\bibinfo {author} {\bibfnamefont {M.}~\bibnamefont
  {de~Gosson}},\ }\bibinfo {title} {On the purity and entropy of mixed
  {G}aussian states},\ in\ \href {https://doi.org/10.1007/978-3-030-05210-2_5}
  {\emph {\bibinfo {booktitle} {Landscapes of Time-Frequency Analysis}}},\
  \bibinfo {editor} {edited by\ \bibinfo {editor} {\bibfnamefont
  {P.}~\bibnamefont {Boggiatto}}, \bibinfo {editor} {\bibfnamefont
  {E.}~\bibnamefont {Cordero}}, \bibinfo {editor} {\bibfnamefont
  {M.}~\bibnamefont {de~Gosson}}, \bibinfo {editor} {\bibfnamefont {H.~G.}\
  \bibnamefont {Feichtinger}}, \bibinfo {editor} {\bibfnamefont
  {F.}~\bibnamefont {Nicola}}, \bibinfo {editor} {\bibfnamefont
  {A.}~\bibnamefont {Oliaro}},\ and\ \bibinfo {editor} {\bibfnamefont
  {A.}~\bibnamefont {Tabacco}}}\ (\bibinfo  {publisher} {Springer International
  Publishing},\ \bibinfo {address} {Cham, Germany},\ \bibinfo {year} {2019})\
  pp.\ \bibinfo {pages} {145--158}\BibitemShut {NoStop}%
\bibitem [{\citenamefont {Agarwal}(1971)}]{Agarwal1971}%
  \BibitemOpen
  \bibfield  {author} {\bibinfo {author} {\bibfnamefont {G.~S.}\ \bibnamefont
  {Agarwal}},\ }\bibfield  {title} {\bibinfo {title} {Entropy, the {W}igner
  distribution function, and the approach to equilibrium of a system of coupled
  harmonic oscillators},\ }\href {https://doi.org/10.1103/PhysRevA.3.828}
  {\bibfield  {journal} {\bibinfo  {journal} {Phys. Rev. A}\ }\textbf {\bibinfo
  {volume} {3}},\ \bibinfo {pages} {828} (\bibinfo {year} {1971})}\BibitemShut
  {NoStop}%
\bibitem [{\citenamefont {Holevo}\ \emph {et~al.}(1999)\citenamefont {Holevo},
  \citenamefont {Sohma},\ and\ \citenamefont {Hirota}}]{Holevo1999}%
  \BibitemOpen
  \bibfield  {author} {\bibinfo {author} {\bibfnamefont {A.~S.}\ \bibnamefont
  {Holevo}}, \bibinfo {author} {\bibfnamefont {M.}~\bibnamefont {Sohma}},\ and\
  \bibinfo {author} {\bibfnamefont {O.}~\bibnamefont {Hirota}},\ }\bibfield
  {title} {\bibinfo {title} {Capacity of quantum {G}aussian channels},\ }\href
  {https://doi.org/10.1103/PhysRevA.59.1820} {\bibfield  {journal} {\bibinfo
  {journal} {Phys. Rev. A}\ }\textbf {\bibinfo {volume} {59}},\ \bibinfo
  {pages} {1820} (\bibinfo {year} {1999})}\BibitemShut {NoStop}%
\bibitem [{\citenamefont {Demarie}(2018)}]{Demarie2018}%
  \BibitemOpen
  \bibfield  {author} {\bibinfo {author} {\bibfnamefont {T.~F.}\ \bibnamefont
  {Demarie}},\ }\bibfield  {title} {\bibinfo {title} {Pedagogical introduction
  to the entropy of entanglement for {G}aussian states},\ }\href
  {https://doi.org/10.1088/1361-6404/aaaad0} {\bibfield  {journal} {\bibinfo
  {journal} {European Journal of Physics}\ }\textbf {\bibinfo {volume} {39}},\
  \bibinfo {pages} {035302} (\bibinfo {year} {2018})}\BibitemShut {NoStop}%
\bibitem [{\citenamefont {Simon}(2000)}]{Simon2000}%
  \BibitemOpen
  \bibfield  {author} {\bibinfo {author} {\bibfnamefont {R.}~\bibnamefont
  {Simon}},\ }\bibfield  {title} {\bibinfo {title} {Peres-{H}orodecki
  {S}eparability {C}riterion for {C}ontinuous {V}ariable {S}ystems},\ }\href
  {https://doi.org/10.1103/PhysRevLett.84.2726} {\bibfield  {journal} {\bibinfo
   {journal} {Phys. Rev. Lett.}\ }\textbf {\bibinfo {volume} {84}},\ \bibinfo
  {pages} {2726} (\bibinfo {year} {2000})}\BibitemShut {NoStop}%
\bibitem [{\citenamefont {Duan}\ \emph {et~al.}(2000)\citenamefont {Duan},
  \citenamefont {Giedke}, \citenamefont {Cirac},\ and\ \citenamefont
  {Zoller}}]{Duan2000}%
  \BibitemOpen
  \bibfield  {author} {\bibinfo {author} {\bibfnamefont {L.-M.}\ \bibnamefont
  {Duan}}, \bibinfo {author} {\bibfnamefont {G.}~\bibnamefont {Giedke}},
  \bibinfo {author} {\bibfnamefont {J.~I.}\ \bibnamefont {Cirac}},\ and\
  \bibinfo {author} {\bibfnamefont {P.}~\bibnamefont {Zoller}},\ }\bibfield
  {title} {\bibinfo {title} {Inseparability {C}riterion for {C}ontinuous
  {V}ariable {S}ystems},\ }\href {https://doi.org/10.1103/PhysRevLett.84.2722}
  {\bibfield  {journal} {\bibinfo  {journal} {Phys. Rev. Lett.}\ }\textbf
  {\bibinfo {volume} {84}},\ \bibinfo {pages} {2722} (\bibinfo {year}
  {2000})}\BibitemShut {NoStop}%
\bibitem [{\citenamefont {Werner}\ and\ \citenamefont
  {Wolf}(2001)}]{Werner2001}%
  \BibitemOpen
  \bibfield  {author} {\bibinfo {author} {\bibfnamefont {R.~F.}\ \bibnamefont
  {Werner}}\ and\ \bibinfo {author} {\bibfnamefont {M.~M.}\ \bibnamefont
  {Wolf}},\ }\bibfield  {title} {\bibinfo {title} {Bound {E}ntangled {G}aussian
  {S}tates},\ }\href {https://doi.org/10.1103/PhysRevLett.86.3658} {\bibfield
  {journal} {\bibinfo  {journal} {Phys. Rev. Lett.}\ }\textbf {\bibinfo
  {volume} {86}},\ \bibinfo {pages} {3658} (\bibinfo {year}
  {2001})}\BibitemShut {NoStop}%
\bibitem [{\citenamefont {Giedke}\ \emph {et~al.}(2001)\citenamefont {Giedke},
  \citenamefont {Kraus}, \citenamefont {Lewenstein},\ and\ \citenamefont
  {Cirac}}]{Giedke2001}%
  \BibitemOpen
  \bibfield  {author} {\bibinfo {author} {\bibfnamefont {G.}~\bibnamefont
  {Giedke}}, \bibinfo {author} {\bibfnamefont {B.}~\bibnamefont {Kraus}},
  \bibinfo {author} {\bibfnamefont {M.}~\bibnamefont {Lewenstein}},\ and\
  \bibinfo {author} {\bibfnamefont {J.~I.}\ \bibnamefont {Cirac}},\ }\bibfield
  {title} {\bibinfo {title} {Entanglement {C}riteria for {A}ll {B}ipartite
  {G}aussian {S}tates},\ }\href {https://doi.org/10.1103/PhysRevLett.87.167904}
  {\bibfield  {journal} {\bibinfo  {journal} {Phys. Rev. Lett.}\ }\textbf
  {\bibinfo {volume} {87}},\ \bibinfo {pages} {167904} (\bibinfo {year}
  {2001})}\BibitemShut {NoStop}%
\bibitem [{\citenamefont {Simon}(2003)}]{Simon2003}%
  \BibitemOpen
  \bibfield  {author} {\bibinfo {author} {\bibfnamefont {R.}~\bibnamefont
  {Simon}},\ }\bibinfo {title} {Separability criterion for {G}aussian states},\
  in\ \href {https://doi.org/10.1007/978-94-015-1258-9_14} {\emph {\bibinfo
  {booktitle} {Quantum Information with Continuous Variables}}},\ \bibinfo
  {editor} {edited by\ \bibinfo {editor} {\bibfnamefont {S.~L.}\ \bibnamefont
  {Braunstein}}\ and\ \bibinfo {editor} {\bibfnamefont {A.~K.}\ \bibnamefont
  {Pati}}}\ (\bibinfo  {publisher} {Springer},\ \bibinfo {address} {Dordrecht,
  The Netherland},\ \bibinfo {year} {2003})\ pp.\ \bibinfo {pages}
  {155--172}\BibitemShut {NoStop}%
\bibitem [{\citenamefont {Jaeger}(2007)}]{Jaeger2007}%
  \BibitemOpen
  \bibfield  {author} {\bibinfo {author} {\bibfnamefont {G.}~\bibnamefont
  {Jaeger}},\ }\href {https://doi.org/10.1007/978-0-387-36944-0} {\emph
  {\bibinfo {title} {Quantum Information: An Overview}}}\ (\bibinfo
  {publisher} {Springer New York},\ \bibinfo {year} {2007})\BibitemShut
  {NoStop}%
\bibitem [{\citenamefont {Makarov}(2018)}]{Makarov2018}%
  \BibitemOpen
  \bibfield  {author} {\bibinfo {author} {\bibfnamefont {D.~N.}\ \bibnamefont
  {Makarov}},\ }\bibfield  {title} {\bibinfo {title} {Coupled harmonic
  oscillators and their quantum entanglement},\ }\href
  {https://doi.org/10.1103/PhysRevE.97.042203} {\bibfield  {journal} {\bibinfo
  {journal} {Phys. Rev. E}\ }\textbf {\bibinfo {volume} {97}},\ \bibinfo
  {pages} {042203} (\bibinfo {year} {2018})}\BibitemShut {NoStop}%
\bibitem [{\citenamefont {Han}\ \emph {et~al.}(1999)\citenamefont {Han},
  \citenamefont {Kim},\ and\ \citenamefont {Noz}}]{Han99}%
  \BibitemOpen
  \bibfield  {author} {\bibinfo {author} {\bibfnamefont {D.}~\bibnamefont
  {Han}}, \bibinfo {author} {\bibfnamefont {Y.~S.}\ \bibnamefont {Kim}},\ and\
  \bibinfo {author} {\bibfnamefont {M.~E.}\ \bibnamefont {Noz}},\ }\bibfield
  {title} {\bibinfo {title} {Illustrative example of {F}eynman’s rest of the
  universe},\ }\href {https://doi.org/10.1119/1.19192} {\bibfield  {journal}
  {\bibinfo  {journal} {American Journal of Physics}\ }\textbf {\bibinfo
  {volume} {67}},\ \bibinfo {pages} {61} (\bibinfo {year} {1999})}\BibitemShut
  {NoStop}%
\bibitem [{\citenamefont {D\'{\i}az}\ \emph {et~al.}(2022)\citenamefont
  {D\'{\i}az}, \citenamefont {Gonz\'alez}, \citenamefont {J.~Hern\'andez},\
  and\ \citenamefont {Vergara}}]{dggv21}%
  \BibitemOpen
  \bibfield  {author} {\bibinfo {author} {\bibfnamefont {B.}~\bibnamefont
  {D\'{\i}az}}, \bibinfo {author} {\bibfnamefont {D.}~\bibnamefont
  {Gonz\'alez}}, \bibinfo {author} {\bibfnamefont {M.}~\bibnamefont
  {J.~Hern\'andez}},\ and\ \bibinfo {author} {\bibfnamefont {J.~D.}\
  \bibnamefont {Vergara}},\ }\bibfield  {title} {\bibinfo {title} {Classical
  analogs of generalized purities and entropies},\ }\href@noop {} {\bibfield
  {journal} {\bibinfo  {journal} {In preparation}\ } (\bibinfo {year}
  {2022})}\BibitemShut {NoStop}%
\bibitem [{\citenamefont {Santhanam}\ \emph {et~al.}(2008)\citenamefont
  {Santhanam}, \citenamefont {Sheorey},\ and\ \citenamefont
  {Lakshminarayan}}]{Santhanam2008}%
  \BibitemOpen
  \bibfield  {author} {\bibinfo {author} {\bibfnamefont {M.~S.}\ \bibnamefont
  {Santhanam}}, \bibinfo {author} {\bibfnamefont {V.~B.}\ \bibnamefont
  {Sheorey}},\ and\ \bibinfo {author} {\bibfnamefont {A.}~\bibnamefont
  {Lakshminarayan}},\ }\bibfield  {title} {\bibinfo {title} {Effect of
  classical bifurcations on the quantum entanglement of two coupled quartic
  oscillators},\ }\href {https://doi.org/10.1103/PhysRevE.77.026213} {\bibfield
   {journal} {\bibinfo  {journal} {Phys. Rev. E}\ }\textbf {\bibinfo {volume}
  {77}},\ \bibinfo {pages} {026213} (\bibinfo {year} {2008})}\BibitemShut
  {NoStop}%
\bibitem [{\citenamefont {Bianchi}\ \emph {et~al.}(2018)\citenamefont
  {Bianchi}, \citenamefont {Hackl},\ and\ \citenamefont
  {Yokomizo}}]{Bianchi2018}%
  \BibitemOpen
  \bibfield  {author} {\bibinfo {author} {\bibfnamefont {E.}~\bibnamefont
  {Bianchi}}, \bibinfo {author} {\bibfnamefont {L.}~\bibnamefont {Hackl}},\
  and\ \bibinfo {author} {\bibfnamefont {N.}~\bibnamefont {Yokomizo}},\
  }\bibfield  {title} {\bibinfo {title} {Linear growth of the entanglement
  entropy and the {K}olmogorov-{S}inai rate},\ }\href
  {https://doi.org/10.1007/JHEP03(2018)025} {\bibfield  {journal} {\bibinfo
  {journal} {Journal of High Energy Physics}\ }\textbf {\bibinfo {volume}
  {2018}},\ \bibinfo {pages} {25} (\bibinfo {year} {2018})}\BibitemShut
  {NoStop}%
\bibitem [{\citenamefont {Lerose}\ and\ \citenamefont
  {Pappalardi}(2020)}]{Silvia2020}%
  \BibitemOpen
  \bibfield  {author} {\bibinfo {author} {\bibfnamefont {A.}~\bibnamefont
  {Lerose}}\ and\ \bibinfo {author} {\bibfnamefont {S.}~\bibnamefont
  {Pappalardi}},\ }\bibfield  {title} {\bibinfo {title} {Bridging entanglement
  dynamics and chaos in semiclassical systems},\ }\href
  {https://doi.org/10.1103/PhysRevA.102.032404} {\bibfield  {journal} {\bibinfo
   {journal} {Phys. Rev. A}\ }\textbf {\bibinfo {volume} {102}},\ \bibinfo
  {pages} {032404} (\bibinfo {year} {2020})}\BibitemShut {NoStop}%
\bibitem [{\citenamefont {Robinett}(1995)}]{Robinett}%
  \BibitemOpen
  \bibfield  {author} {\bibinfo {author} {\bibfnamefont {R.~W.}\ \bibnamefont
  {Robinett}},\ }\bibfield  {title} {\bibinfo {title} {Quantum and classical
  probability distributions for position and momentum},\ }\href
  {https://doi.org/10.1119/1.17807} {\bibfield  {journal} {\bibinfo  {journal}
  {American Journal of Physics}\ }\textbf {\bibinfo {volume} {63}},\ \bibinfo
  {pages} {823} (\bibinfo {year} {1995})}\BibitemShut {NoStop}%
\bibitem [{\citenamefont {Gattus}\ and\ \citenamefont
  {Karamitsos}(2020)}]{Gattus}%
  \BibitemOpen
  \bibfield  {author} {\bibinfo {author} {\bibfnamefont {V.}~\bibnamefont
  {Gattus}}\ and\ \bibinfo {author} {\bibfnamefont {S.}~\bibnamefont
  {Karamitsos}},\ }\bibfield  {title} {\bibinfo {title} {Dimensional analysis
  and the correspondence between classical and quantum uncertainty},\ }\href
  {https://doi.org/10.1088/1361-6404/aba6bc} {\bibfield  {journal} {\bibinfo
  {journal} {European Journal of Physics}\ }\textbf {\bibinfo {volume} {41}},\
  \bibinfo {pages} {065407} (\bibinfo {year} {2020})}\BibitemShut {NoStop}%
\end{thebibliography}%

\end{document}